\documentclass[ amsmath, amssymb, aps, prb, twocolumn, superscriptaddress, floatfix ]{revtex4-1}%
\usepackage{graphicx} 
\usepackage{siunitx}
\usepackage{dcolumn} 
\usepackage{bm} 
\usepackage{xspace} 
\usepackage{placeins} 
\usepackage{tikz} 
\usepackage{amsmath}
\usepackage{natbib}
\usepackage{amsfonts}
\usepackage{amssymb}
\usepackage{upgreek}
\usepackage{textcomp}

\providecommand{\U}[1]{\protect\rule{.1in}{.1in}}

\setcitestyle{super}

\begin{document}
\title{Superconductivity in La$_{2}$Ni$_{2}$In}
\author{Jannis Maiwald}
\affiliation{Quantum Matter Institute, University of British Columbia, Vancouver, BC, Canada}
\affiliation{Department of Physics and Astronomy, University of British Columbia, Vancouver, BC, Canada}
\email{jannis.maiwald@ubc.ca}
 \author{Igor I. Mazin}
 \affiliation{Department of Physics and Astronomy, George Mason University, Fairfax, VA, USA }
 \affiliation{Quantum Science and Engineering Center, George Mason University, Fairfax, VA,
 USA. }
 \author{Alex Gurevich}
 \affiliation{Department of Physics, Old Dominion University, Norfolk, Virginia 23529, USA
 }
\author{Meigan Aronson}
\affiliation{Quantum Matter Institute, University of British Columbia, Vancouver, BC, Canada}
\affiliation{Department of Physics and Astronomy, University of British Columbia, Vancouver, BC, Canada}

\date{\today}

\begin{abstract}
We report here the properties of single crystals of La$_2$Ni$_2$In. Electrical resistivity and specific heat measurements concur with the results of Density Functional Theory (DFT) calculations, finding that La$_{2}$Ni$_{2}$In is a weakly correlated metal, where the Ni magnetism is almost completely quenched, leaving only a weak Stoner enhancement of the density of states. Superconductivity is observed at temperatures below \SI{0.9}{\kelvin}. A detailed analysis of the field and temperature dependencies of the resistivity, magnetic susceptibility, and specific heat at the lowest temperatures reveals that La$_2$Ni$_2$In is a dirty type-II superconductor with likely s-wave gap symmetry. Nanoclusters of ferromagnetic inclusions significantly affect the subgap states resulting in a non-exponential temperature dependence of the specific heat $C(T)$ at $T\ll T_\text{c}$.


\end{abstract}

\pacs{Valid PACS appear here}
\maketitle

\sisetup{range-phrase=--}

\section{Introduction}

Electronic correlations lead to an array of different ordered and disordered phases in metals. These phases are especially interesting when they occur on low dimensional or geometrically frustrated lattices. If the correlations are not strong enough to render the system fully insulating, the stabilization of increasingly localized electronic states and the associated magnetic moments adds a compelling complexity to the phase behaviors of correlated metals. Finding isostructural, and even isoelectronic, systems where the relationships between the different phases and behaviors can be explored in a controlled way is of great importance. The $R_{2}T_{2}X$ ($R$=rare earth, $T$=transition metal, $X$=main group, pnictogen, chalcogen) family has proven particularly useful, due to its unique crystal structure~\cite{Rodewald2007,Gschneidnder2004,Lukachuk2003}. The lattice of f-electron bearing rare earth atoms $R$ consists of planes of orthogonal dimers, as in the Shastry-Sutherland lattice (SSL), that are stacked along the c-axis. Depending on the relative distances between the rare earth moments in the plane or perpendicular to the plane, two dimensional systems of isolated SSL planes~\cite{kraft2004,kim2011} can be realized, or alternatively spin ladder systems~\cite{Wu2016,Gannon2019}. For $R_{2}T_{2}X$ ($R$=Ce,Yb), the f-electrons of the rare earths can also display mixed valence behaviors~\cite{Gordon1995,sereni2010,gannon2018,Kaczorowski1996,Giovannini1998,Dhar2007}, indicating that coupling of the f-electrons to extended states can lead to the overall suppression of magnetic moments. It is expected that the interplay of strong quantum fluctuations due to the unique $R_{2}T_{2}X$ lattice\cite{Dhar2007,Hauser1997,Havela1994}, and as well proximity to the delocalization of the f-electrons via the Kondo effect can result in the destabilization of ordered phases, and to the formation of quantum critical points (QCPs) where the host systems are transitioning among different phases~\cite{sereni2010,muramatsu2011,lamura2020,Dhar2007,Bauer2005}. Given this richness of behavior when the rare earth $R$ has f-electrons, it is surprising that relatively little is known about compounds from this family where $R$=La,Lu,Y , where there are no valence f-electrons. Here, there is the possibility of instead studying d-electron magnetism and correlations that are associated with the $T$ atoms. At present, 21 compounds have been identified where $R$=Y,La,Lu, $T$=Ni,Cu,Au,Pt,Pd,Rh, and $X$=Mg,In,Pb,Cd. However, there has been relatively little study of the physical properties of these compounds~\cite{Kaczorowski1996}, and to our knowledge none of these studies have been carried out on single crystals. 

Of the known $R_{2}T_{2}X$ compounds with $R$=Y,La,Lu, it seems likely that compositions with $T$=Ni have magnetic character. For that reason, we focused initially on La$_{2}$Ni$_{2}$In. Previous measurements were carried out on polycrystalline samples~\cite{Kaczorowski1996}, where the possibility of small inclusions of multiple phases, and the lack of a well defined conduction path could potentially complicate interpretation. We report here the first detailed study of the basic properties of La$_{2}$Ni$_{2}$In crystals. Electrical resistivity measurements find that this compound is metallic. Density Functional Theory (DFT) calculations of the electronic structure reveal that there is considerable charge transfer associated with the Ni d-states, which are well below the Fermi level. Thus, the manifestations of the d-electron character are very weak. Specific heat measurements concur that the mass enhancement of the quasiparticles is quite small, in good agreement with the DFT calculations. Measurements of the magnetic susceptibility confirm the weak magnetism predicted by the DFT calculations, revealing that there are no localized magnetic moments, and the only intrinsic part of the magnetic susceptibility is the temperature-independent Pauli susceptibility with a minimal Stoner enhancement. Overall, La$_{2}$Ni$_{2}$In is a weakly correlated metal, with minimal magnetic character. We find that it undergoes a transition to a superconducting state below 0.89\,K. A detailed analysis of the specific heat, electrical resistivity, and magnetic susceptibility in the superconducting state was carried out, revealing that La$_{2}$Ni$_{2}$In is a conventional Type-II superconductor in the dirty limit.

\section{Methods}

Single crystals of La$_{2}$Ni$_{2}$In were synthesized
using a LaNi self-flux. The
precursor LaNi was prepared using solid-state reaction. Stoichiometric amounts
of La (99.9\,\%) and Ni (99.995\,\%) were cut into small pieces and placed
into an Al$_{2}$O$_{3}$ crucible, sealed under $\sim$ 250\,mbar argon
atmosphere in a quartz tube and 
successively heated to about \ang{1100}C for
40\,h. The above steps were carried out under a protective argon atmosphere to
prevent Lanthanum from oxidizing. After synthesis the precursor was ground
into a powder.

The precursor and elemental In (99.999\,\%) were mixed in a ratio of 6:1 and
placed in a Canfield Crucible Set (two Al$_{2}$O$_{3}$ crucibles separated by
a strainer) and then sealed as described above. The sealed quartz tubes were heated in an
open furnace over the course of 3\,h to \ang{770}C, where they remained for 5\,h to ensure adequate mixing of
the reactants. Growth of La$_{2}$Ni$_{2}$In single crystals took place
while lowering the temperature over a period of 15\,h down to \ang{725}C. At
this temperature the still liquid self-flux was separated from the crystals by
placing the batch upside down in a centrifuge and spinning it at 2000-3000\,rpm
for about 15\,s. Using this procedure we were able to grow needle-like single
crystals of La$_{2}$Ni$_{2}$In for the first time. Typical dimensions
of the crystals are about 0.5\,$\times$\,0.5\,$\times$\,4\,mm$^{3}$. An image of the five single crystals used for heat capacity measurements is shown in the Fig.~\ref{fig_xrd}(a).

Powder x-ray
diffraction (XRD) patterns were recorded with a Bruker D8 Focus diffractometer in the Bragg-Bretano configuration using a Co cathode. The program FullProf was used to refine the powder diffraction patterns. Measurements of
the electrical resistivity and specific heat were obtained using a Physical
Property Measurements System (PPMS) from Quantum Design equipped with a He$^{3}$/He$^{4}$
dilution refrigerator insert. Measurements of the magnetic susceptibility were done using a Magnetic Property Measurements System 3 (MPMS3) also from Quantum
Design and equipped with a He$^{3}$ insert.

Band structure calculations were performed using the Linear Augmented Plane
Wave code WIEN2k\cite{Wien2k}. For the calculations of the zone-center phonons
the VASP pseudopotential package was used\cite{Vasp}. The gradient-corrected
density functional of Ref.~\onlinecite{Perdew1996} was used in all calculations.

\section{Normal state properties}

\subsection{Sample characterization}

\begin{table}[b]
\caption{Results of the XRD Rietvelt refinement.}%
\label{tab_refinement}
\begin{ruledtabular}
		\begin{tabular}{llll}
		Lattice Parameters & a [\si{\nano\meter}] & c [\si{\nano\meter}] & c/a \\
		\hline
		& & &\\
		Single Crystal & 0.76448(2) & 0.389149(14) & 0.5009 \vspace{2.5mm} \\
		
		Atomic Coordinates & x & y & z \vspace{1.5mm} \\
		La (4h)	 & 0.17708 & 0.67708 & 0.5\\
		Ni (4g) & 0.61708 & 0.11708 & 0.0 \\
		In (2a) & 0.0 & 0.0 & 0.0 \\
			\end{tabular}
	\end{ruledtabular}
\end{table}

\begin{figure}[ptb]
\centering
\includegraphics[width=0.95\linewidth]{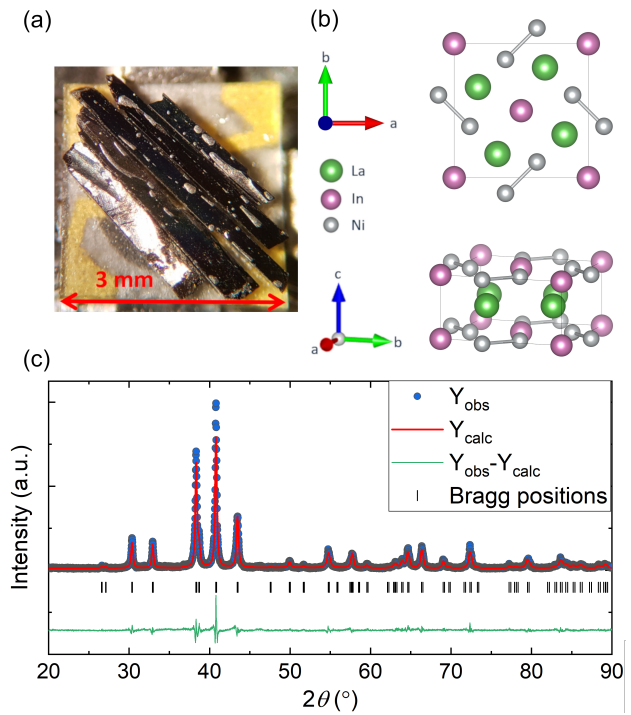}
\par
\caption{(a) Optical microscope image of five single crystals of La$_2$Ni$_2$In. (b) The refined crystal structure of La$_{2}$Ni$_{2}$In. The unit cell (thin grey lines) is shown in (top) top view and (bottom) perspective view. Grey solid lines depict the nearest neighbour Ni distance of \SI{0.253}{\nano\meter}. (c) Measured powder XRD pattern of as grown La$_{2}$Ni$_{2}$In, and its refinement using the reported tetragonal structure with space group P4/mbm. Bragg peak positions are indicated.}%
\label{fig_xrd}%
\end{figure}

The as-grown single crystals of La$_{2}$Ni$_{2}$In were structurally characterized by powder x-ray
diffraction. The diffraction pattern and its refinement are compared in Fig.~\ref{fig_xrd}. All recorded diffraction peaks can be indexed within the
reported tetragonal structure \cite{Lukachuk2003}, which has the space group P4/mbm. The absence of extrinsic diffraction peaks rules out crystalline impurity phases with concentrations larger than $\simeq$\SI{1}{\percent}. The refined lattice parameters and atomic coordinates are presented in Table~\ref{tab_refinement}. In particular, the refined a and c lattice constants for our single crystals lie within the range of values reported for annealed polycrystalline samples\cite{Kalychak1990,Kaczorowski1996}. This indicates that the presence of defects and/or non-stoichiometry in the latter is rather limited. We attribute the minor differences between the observed and refined
intensities in the XRD pattern to possible vacancies and/or site exchange, which are not uncommon
in polycrystalline $R_{2}T_{2}X$ compounds\cite{Pustovoychenko2012}. Finally, we found
no indication of a phase with an orthorhombic structure, as reported earlier\cite{Pustovoychenko2012}. In fact, forcing the orthorhombic structure
for the Rietveld refinement requires additional peaks that are not observed in our measured data. A comparison between the tetragonal and orthorhombic refinements can be found in Appendix~\ref{sec_ortho}.

Like most of the $R_{2}T_{2}X$ compounds, La$_{2}$Ni$_{2}$In adopts the tetragonal Mo$_{2}$FeB$_{2}$ type structure with space group P4/mbm, which is
an ordered derivative of the U$_{3}$Si$_{2}$ structure where the $X$ atom
occupies one of the two inequivalent U sites. Fig.\ref{fig_xrd}(b) shows that, like the $R$ atoms, the Ni atoms form dimers with the closest Ni-Ni distance (solid grey bonds) of \SI{0.253}{\nano\meter}. The dimers form a square lattice in the a-b plane, where the closest interdimer distance is \SI{0.541}{\nano\meter}. The Ni-Ni distance along the c axis is \SI{0.389}{\nano\meter}, indicating a quasi 2-dimensional environment for the Ni subsystem.

\subsection{Density functional calculations of the electronic structure}

\begin{figure*}[ptb]
\centering
\includegraphics[width=\linewidth]{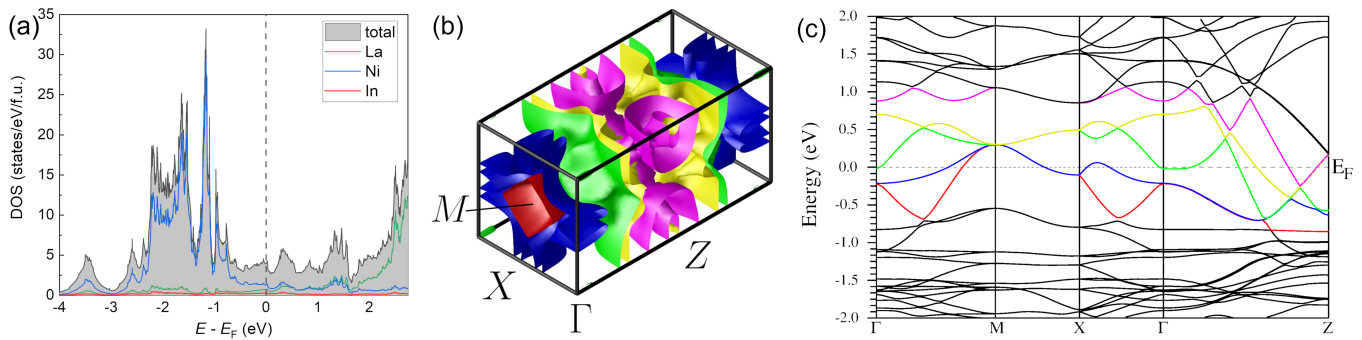}
\caption{Density functional calculation results for La$_2$Ni$_2$In. (a) Density of states as a function of energy $E$ relative to the Fermi energy $E_\text{F}$. (b) Fermi surface calculated using the tetragonal crystal structure determined from our XRD analysis. The colors refer to Fermi level crossings of bands with the same colors (c). (c) Electronic band structure plotted along high symmetry directions.}%
\label{fig_fermi}%
\end{figure*}

We have performed ab initio band structure calculations using the experimentally determined tetragonal structure of La$_{2}$Ni$_{2}$In.
The calculated density of states (DOS) and Fermi surface of La$_2$Ni$_2$In are depicted in Fig.~\ref{fig_fermi}. In view of the above-mentioned proposal of an orthorhombic structural variant~\cite{Pustovoychenko2012}, we have as well
attempted a full structural optimization in both crystallographic groups. The tetragonal P4/mbm structure optimized into $a=0.7663$\,nm and $c= 0.3906$\,nm, with the internal parameters, in the order of Table~\ref{tab_refinement}, being 0.1772, 0.6772, 0.6181, and 0.1181. The orthorhombic Pbam structure optimizes into $a=0.7643$, $b=1.5595$ and $c=0.3903$\,nm, with La 4g (0.6068, 0.2840, 0) and (0.7438, 0.5361, 0), Ni 4h (0.5565, 0.4251, 0.5), and (0.3044, 0.3189, 0.5), In 4h (0.4259, 0.1276, 0.5). Curiously, the latter structure is lower in energy by $\approx 30$\,meV/formula, consistent with the orthorhombic structure found in polycrystalline samples~\cite{Pustovoychenko2012}, but not in our single crystal samples, where the tetragonal structure is stabilized. We will use the results from the tetragonal refinement for our analysis of La$_{2}$Ni$_{2}$In.

The Ni-based states form a broad band between $\simeq$ -2\,eV and -1\,eV (Fig.~\ref{fig_fermi}a). The breadth of the band reflects a combination of hybridization and charge transfer, indicating that Ni likely has little magnetic character. There is a robust density of states at the Fermi level E$_{F}$, confirming that La$_{2}$Ni$_{2}$In is metallic. The Ni d-states contribute little weight at E$_{F}$, and so we conclude that the metal will have at most weak magnetic correlations. Accordingly, we also performed fixed spin moment calculations to determine the Stoner-renormalized spin susceptibility, which appears to be about 33\% higher than the Pauli susceptibility. This is a modest Stoner enhancement, typical for nonmagnetic metals such as Al. While Ni is often magnetic or close to magnetic in many of its compounds, magnetism is almost completely quenched in La$_{2}$Ni$_{2}$In.

The Fermi surface of La$_{2}$Ni$_{2}$In is depicted in Fig.~\ref{fig_fermi}(b). The five sheets of the Fermi surface correspond to band-crossings that are evident in the electronic band structure (Fig.~\ref{fig_fermi}c). It is notable that two of these sheets show little dispersion along $\Gamma$-M, possibly suggesting a weak two-dimensionality that is not inconsistent with the layered character of the La$_{2}$Ni$_{2}$In crystal structure (Fig.~1). The calculated plasma frequencies along the crystallographic a and c directions are \SI{3.31}{\electronvolt} and \SI{3.98}{\electronvolt}, corresponding to the Fermi velocities of $0.19\times 10^8$ and $0.23\times 10^8$ cm/sec, respectively. In addition, we calculated the frequencies of the zone-center phonons, which can be found in Appendix~\ref{appendix_dft}.

\subsection{Specific Heat}

\begin{figure}[ptb]
\centering
\includegraphics[width=\linewidth]{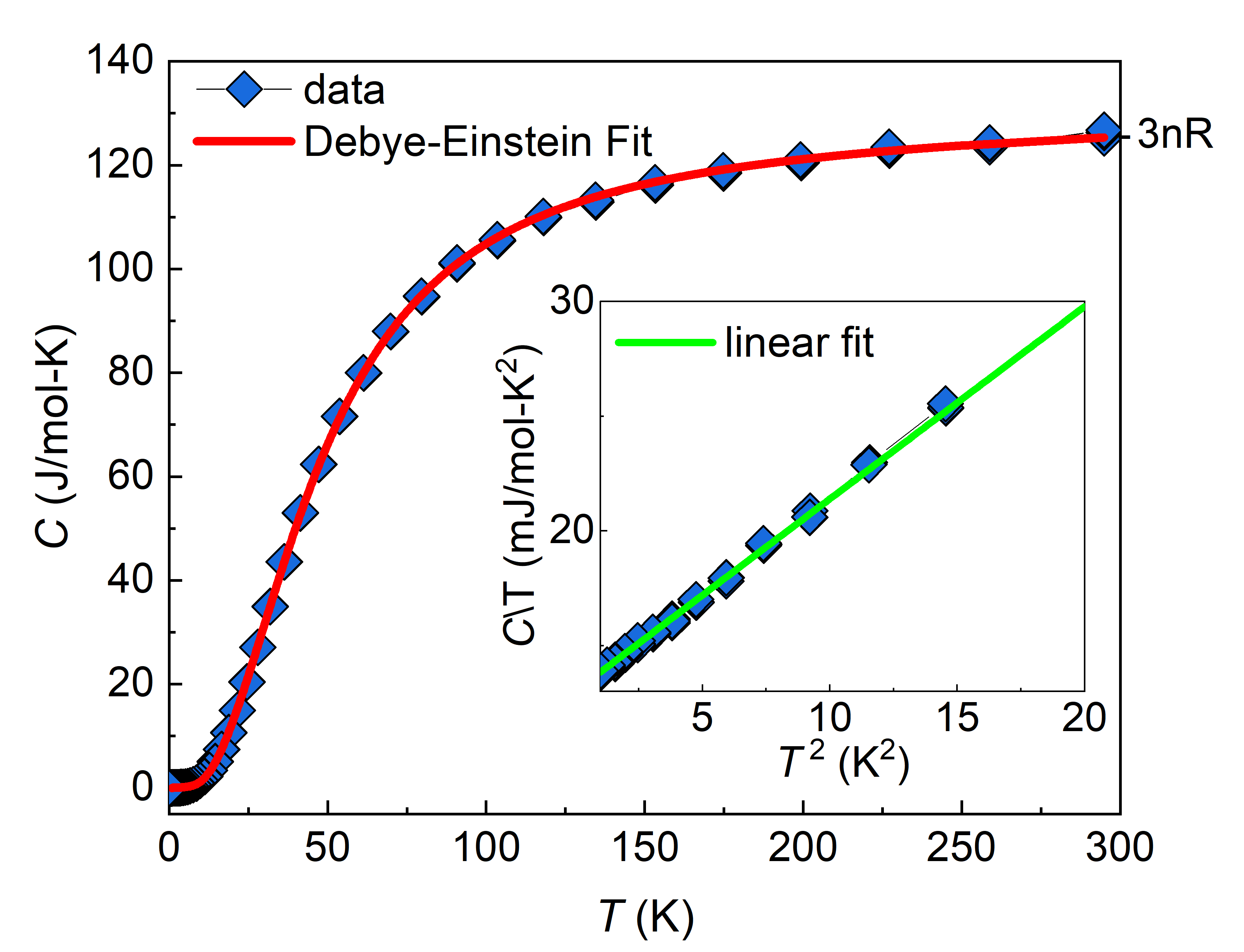}
\caption{Specific heat $C(T)$ of La$_{2}$Ni$_{2}$In as a function of
temperature $T$ between \SIrange{0.05}{300}{\kelvin}. The red line is a fit to
the data using the model defined in the main text. 3$nR$ denotes the
Dulong-Petit constant with n=5. (Inset) Specific
heat $C/T$ as a function of $T^{2}$ between \SIrange{1}{4.5}{\kelvin}. The solid
green line is a fit to the expression: $C(T)=\gamma T+\beta T^{3}$, yielding a
Sommerfeld coefficient of $\gamma
=\SI[mode = text]{13.01\pm0.03}{\milli\joule\per\mole\per\kelvin\tothe{2} }$.}%
\label{fig_hc}%
\end{figure}

The temperature dependence of the specific heat $C(T)$ of La$_{2}$Ni$_{2}$In is presented in Fig.~\ref{fig_hc}. $C$ rises monotonically with increasing temperatures $T$ between \SIrange{0.05}{300}{\kelvin}, reaching the Dulong-Petit
value of $3nR$ with $n=5$ atoms per formula unit (f.u.) around \SI{300}{\kelvin}. However, our data are not well described by the standard Debye model alone. Since the Debye model only accounts for acoustic modes, this may suggest that low-energy optical modes may be present in La$_{2}$Ni$_{2}$In. This expectation was confirmed in our calculations of the zone-center phonon spectrum (see Table~\ref{tab:phonons}, Appendix~\ref{appendix_dft}), where a number of low-lying optical modes were found. This motivated us to model the measured specific heat $C$ using a Debye-Einstein model:
\begin{equation}
C=C_{\text{e}}\ +m\cdot C_{\text{D}}\ +(1-m)\cdot C_{\text{E}},
\end{equation}
with weighing factor $m$. The electronic specific heat is given by
$C_{\text{e}}\ =\gamma T$ and the Debye and Einstein terms by
\begin{align*}
C_{\text{D}}\  &  =9nN_{\text{A}}k_{\text{B}}\ \left( \frac{T}{T_{\text{D}}%
}\right)^{3}\int_{0}^{\frac{T_{\text{D}}}{T}}\frac{x^{4}}{(e^{x}%
-1)(1-e^{-x})}dx,\\
C_{\text{E}}\  &  =3nN_{\text{A}}\ k_{\text{B}}\ \left( \frac{T_{\text{E}}%
}{T}\right)^{2}\frac{1}{(e^{T_{\mathrm{E}}{}/T}-1)(1-e^{-T_{\mathrm{E}}{}%
/T})},
\end{align*}
respectively, with the Sommerfeld coefficient $\gamma$, Avogadro's constant $N_{\text{A}}$,
Boltzmann's constant $k_{\text{B}}$ and $T_{\text{E}}$ the Einstein
temperature. The data are well described with the following parameters:
$\gamma=\SI{11.3\pm0.3}{\milli\joule\per\mole\kelvin\tothe{2}}$,
$T_{\text{D}}\ =\SI{211\pm1}{\kelvin}$, $T_{\text{E}}\ =\SI{85\pm1}{\kelvin}$
and $m=\SI{0.82\pm0.01}{}$.

We also extracted $\gamma$ and $T_{\text{D}}$ from a linear fit to $C/T$ as a
function of $T^{2}$ at low temperatures between \SIrange{1}{4}{\kelvin} (see inset of
Fig.~\ref{fig_hc}) using the expression: $C(T)=\gamma T+\beta T^{3}$, with $\beta=12\pi
^{4}/5\cdot nN_{\text{A}}k_{\text{B}}/T_{\text{D}}^{3}$. This procedure resulted in $\gamma
=\SI{13.01\pm0.03}{\milli\joule\per\mole\per\kelvin\tothe{2}}$ and $T_{\text{D}%
}\ =\SI{226\pm1}{\kelvin}$, which are in reasonable agreement with the values found in the more comprehensive fit above. In the following we will use the values
obtained from the low temperature fit. The derived Sommerfeld coefficient
corresponds to \SI{5.52\pm0.03}{states\per\electronvolt\per \text{f.u.}} at
the Fermi level, which is 38\,\% larger than the value from our DFT calculations
($\approx\SI{4.0}{states\per\electronvolt\per \text{f.u.}}$). Given the apparent absence of magnetic correlations in the DFT calculations, we conclude that this mass
renormalization is a consequence of electron-phonon coupling, with a magnitude
$\lambda_{m^\ast/m}=0.38$.

\subsection{Electrical Resistivity}

\begin{figure}[ptb]
\centering
\includegraphics[width=\linewidth]{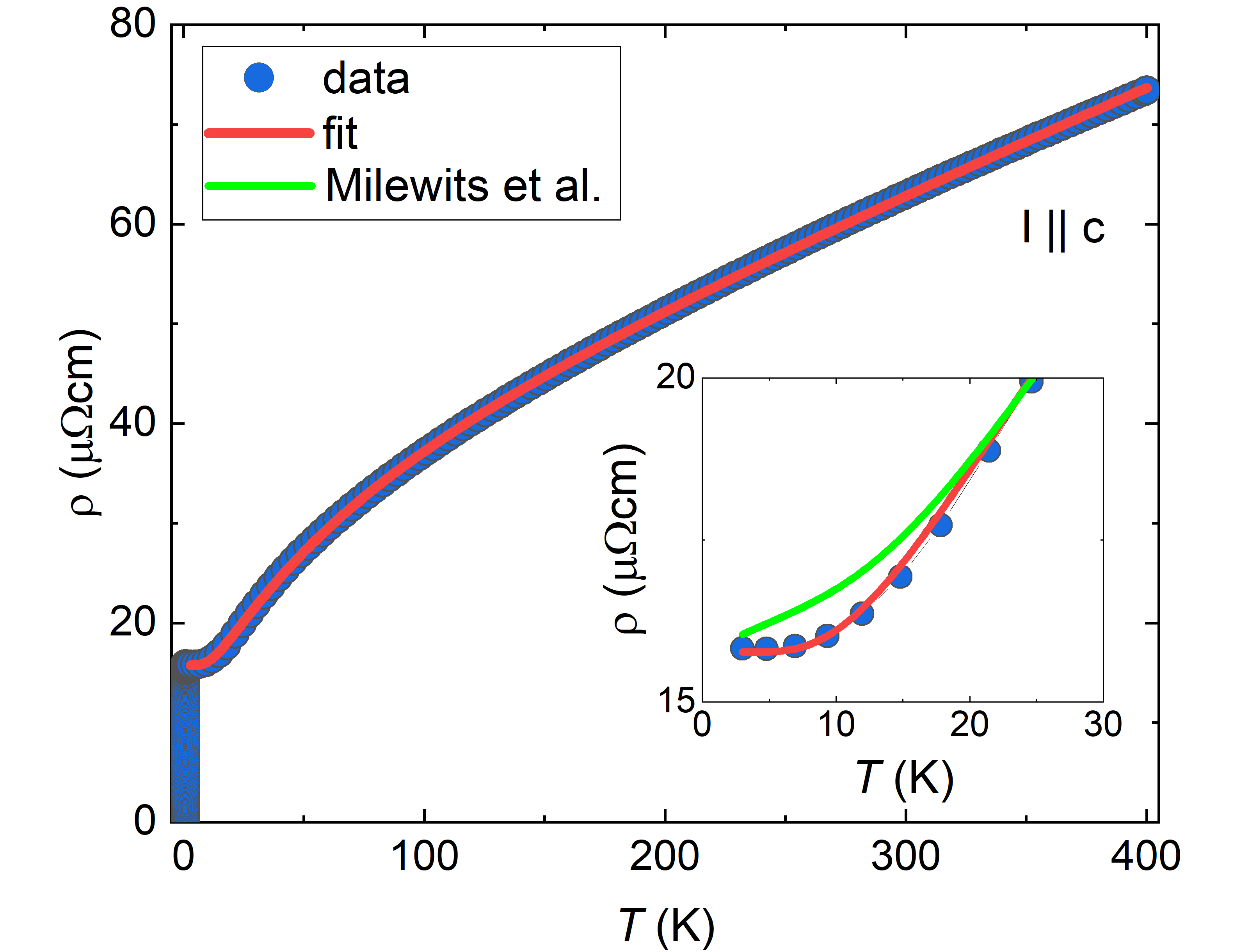}
\caption{Temperature dependence of the electrical resistivity $\rho(T)$ of La$_{2}$Ni$_{2}$In, with the measuring current $I$ applied along the crystallographic c axis. The data (blue symbols) are well described by the heuristic model (red line) with an exponential factor as discussed in the main text. (Insert) Low temperature region of $\rho(T)$, showing the difference between our heurisitc fit and the fit to the model reported by Milewits et al.\cite{Milewits1976} using an exponential term (green line).}
\label{fig_res}
\end{figure}

Fig.~\ref{fig_res} shows the electrical resistivity $\rho(T)$ of La$_{2}%
$Ni$_{2}$In as a function of temperature $T$ in the range
of \SIrange{0.15}{400}{\kelvin}. The measuring current was applied along the
crystallographic c axis. The room-temperature value $\rho_\text{300\,K} = \SI{63}{\mu\Omega\centi\meter}$ identifies La$_{2}$Ni$_{2}$In as a good metal, while the residual resistivity ratio (RRR) $\rho_{\SI{300}{\kelvin}}/\rho_{\SI{2}{\kelvin}}=4$ indicates significant
defect scattering at low temperatures that is evidenced by the rather large residual resistivity $\rho_{0}\simeq \SI{16}{\mu\Omega\centi\meter}$ (inset Fig.~\ref{fig_res}).
Nonetheless, the absolute values of $\rho_{0}$ in our single crystals are smaller than those reported in polycrystalline samples by factors of 3-5~\cite{Kaczorowski1996}.
While this may reflect a higher defect concentration in the polycrystalline samples, we have also noticed the formation of a low-conductivity passivated surface layer on our single crystals when they are exposed to air over time. The greater surface area of polycrystalline samples, combined with the uncertain current paths, could plausibly lead to significantly higher resistivity in polycrystalline samples than in freshly prepared single crystals.

The resistivity of single crystals decreases monotonically with decreasing temperature, as expected in a good metal (Fig.~\ref{fig_res}). The decrease at high temperatures initially has a sublinear slope, leading to an inflection point around \SI{30}{\kelvin}. The resistivity approaches its residual value $\rho_{0}$, but below \SI{0.89}{\kelvin} the resistance drops sharply to zero, indicating a transition into a superconducting state. This transition is extremely sharp, having a width $\Delta T_{\text{c}}\ \approx\SI{80}{\milli\kelvin}$.

The normal state resistivity $\rho(T)$ in the range \SIrange{2}{30}{\kelvin} is reasonably well
described by scattering of weakly correlated quasiparticles from acoustic phonons, as described by the Bloch-Gr\"{u}neisen law:
\begin{equation}
\rho(T)= \rho_{0}+\rho_{\mathrm{D}}(T), \label{BG}
\end{equation}
\begin{equation}
\rho_{\mathrm{D}}
(T)=A\left(\frac{T}{T_{\mathrm{D}}}\right)^{5}\int_{0}^{T_{\mathrm{D}}/T}\frac{x^{5}dx}{\left(e^{x}
-1\right)\left(1-e^{-x}\right)},
\label{rD}
\end{equation}
where $T_{\mathrm{D}}$=226~K is the Debye temperature taken from the specific heat analysis. This fit agrees reasonably well with the measured $\rho$(T) between \SIrange{0.89}{30}{\kelvin}, but not above \SI{30}{\kelvin}.
Inspired by the analysis of the specific heat, we added an Einstein term, $\rho
_{E}(T)=B(T_{\mathrm{E}}/T)\ /(e^{T_{\mathrm{E}}{}/T}-1)(1-e^{-T_{\mathrm{E}%
}{}/T}),$ to the fit. However, this did not lead to any appreciable improvement in the goodness of fit.

Adding an exponential term, $B\exp(-T_{0}/T),$ that accounts for Umklapp processes
assisted by a specific phonon with an energy $T_{0}$~\cite{Woodard1964,Milewits1976}, improves the agreement between the fit and the data,
but does not make it perfect (see green line in Fig.~\ref{fig_res}, inset). Curiously, adding an exponential \textit{factor}, as in
\begin{equation}
\rho(T)=\left(\rho_{0}+\rho_{D}(T)\right) \left[ 1+c\cdot\exp(-T_{0}/T)\right] ,\label{eq_res}%
\end{equation}
generates an essentially perfect fit to the data (Fig.~\ref{fig_res}), with
the following parameters: $\rho_{0}=\SI{15.7\pm0.1}{\mu\Omega\centi\meter}$,
$T_{\mathrm{D}}=\SI{226}{\kelvin}$ (taken from the Debye-Einstein fit to our
specific heat data), $A=\SI{36.1\pm0.2}{\mu\Omega\centi\meter}$,
$c=\SI{1.49\pm0.01}{}$ and $T_{0}=\SI{42.4\pm0.4}{\kelvin}$. It is interesting that T$_{0}$ is essentially half the value of the Einstein mode temperature T$_{E}$=85\,K that was determined from the fit to the specific heat. However, we cannot offer a microscopic explanation for the origin of this exponential factor. 

This fit leads to the linear coefficient $d\rho(T)/dT= \SI{0.10}{\mu\Omega\centi\meter\per\kelvin}$ at large $T$. Using the Drude formula for the phonon-limited resistivity, we find $\lambda_{\text{tr}}%
\ =(d\rho/dT)\omega_{\text{p}}^{2}/4.01$, where $(d\rho/dT)$ is the high-temperature resistivity coefficient in units \si{\mu\Omega\centi\meter\per\kelvin}. Taking the calculated plasma
frequency along the c axis $\omega_{\text{p}}\ =\SI{3.98}{\electronvolt}$, we deduce the electron-phonon coupling constant $\lambda_{\text{tr}}$=0.395, in good agreement with the value $\lambda_{m^\ast/m}=0.38$ from the DFT calculations.

\subsection{Magnetic Susceptibility}

\begin{figure}[ptb]
\centering
\includegraphics[width=\linewidth]{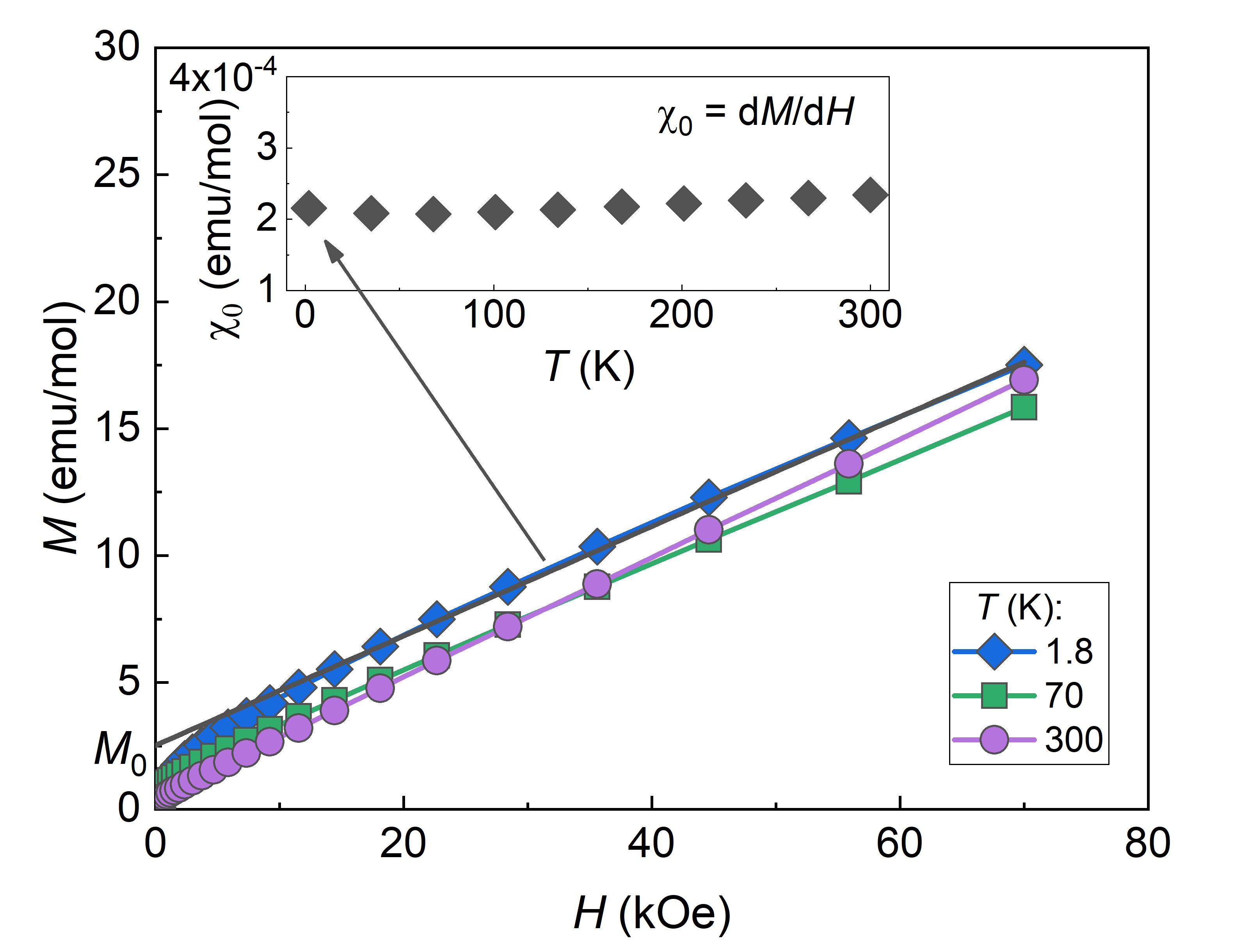}
\caption{Magnetization $M(H)$ of La$_{2}$Ni$_{2}$In as a function of magnetic field $H$ for various selected temperatures $T$, as indicated and the effective susceptibility $\chi_{0}$ (black filled diamonds), determined from $M(H)$ as described in the text. Measurements were carried out on a stack of 6 co-aligned crystals, and the magnetic field is applied parallel to the crystallographic c axis. Extrapolating the high field data to $H=0$ (dashed line) gives the indicated intercept $M_0$, and the slope $\chi_{0}$.}
\label{fig_susT}%
\end{figure}

Although our DFT calculations indicate that La$_2$Ni$_2$In is not appreciably magnetic, measurements of the magnetic moment as a function of the magnetic field
applied along the crystallographic c axis at various temperatures (Fig.~\ref{fig_susT}) reveal nonlinearities at low fields that are suggestive of weak ferromagnetism. Hysteresis is observed in $M(H)$ at low temperatures, with a magnitude that decreases with increasing temperature and the dc magnetic susceptibility $\chi=M/H$ is strongly temperature dependent when measured in low fields, but this temperature dependence weakens appreciably in high fields (see Appendix~\ref{appendix_sample_depend}, Fig.~\ref{fig_supp_sus}).

\begin{figure}[ptb]
\centering
\includegraphics[width=\linewidth]{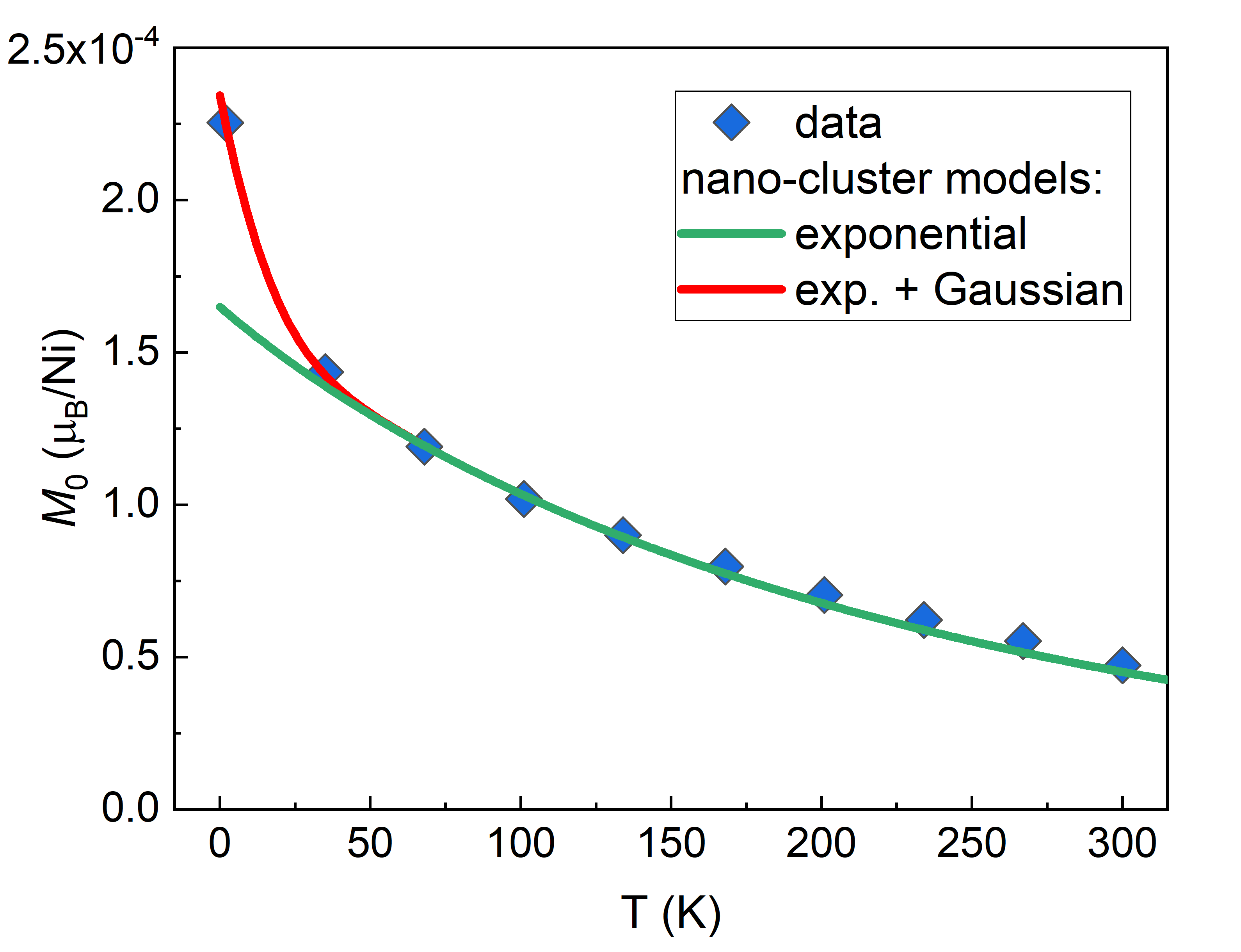}
\caption{
The temperature dependence of the spontaneous magnetization $M_0$ (blue diamonds). The data are compared to two models of the ferromagnetic nanoclusters: exponential distribution of cluster sizes (red line) and exponential distribution with additional Gaussian distribution to describe ultra-small clusters (green line). See Eq.~\eqref{eq_m0} and discussion in the text. Measurements of different samples (Appendix~\ref{appendix_sample_depend}) find that the distribution of nanocluster sizes does not change appreciably among crystals taken from the same batch, although the total amount of magnetic material does differ moderately.} 
\label{fig_mag}%
\end{figure}

The magnetization $M(H)$ isotherms that are presented in Fig.~\ref{fig_susT} are illuminating about the nature of the ferromagnetism that is observed in La$_{2}$Ni$_{2}$In. In the spirit of the Arrott plot\cite{Arrott1957}, $M(H)$ is linear at high fields, and the spontaneous moment $M_{0}$ can be estimated by extrapolating to zero field. The temperature dependence of $M_{0}$ is plotted in Fig.~\ref{fig_mag}, showing that $M_{0}$ decreases slowly from its value of $\sim$ 2.5$\times10^{-4}$\,$\mu_\text{B}/$Ni at 1.8\,K to $\sim$ 0.5$\times10^{-4}$\,$\mu_\text{B}/$Ni at 300\,K. These data indicate that the onset of ferromagnetism occurs well above 300\,K, and the very slow development of $M_{0}$ and the lack of its saturation well below the apparent Curie temperature $T_\text{C}$ is inconsistent with the order-parameter behavior observed in pristine, bulk ferromagnets. The most likely interpretation of these data is that the ferromagnetism originates with a contaminant phase that was introduced during the synthesis on the surface or as an inclusion in our single crystals. Elemental Ni with $T_\text{C}=627$\,K that was not completely reacted in the initial LaNi precursor seems a likely possibility. The measured value $M_{0}\sim$ 2.5$\times10^{-4} \mu_\text{B}/$Ni in La$_{2}$Ni$_{2}$In could be explained by the presence of less than 0.04$\%$ of elemental Ni, with $M_0(\text{Ni})$ $\sim 0.63\,\mu_\text{B}$/Ni\cite{Chiffey1971}. This is far less than could be detected by XRD or most analytical methods. 

We find that the temperature dependence of $M_{0}$ in Fig.~\ref{fig_mag} can be explained quantitatively by assuming the presence of nanoscale clusters, each containing a few
to a few dozens of magnetic ions.
If the Curie temperatures of the individual nanoclusters are distributed by a function
$F(T_\text{C}),$ that is, the probability to find a ferromagnetic Ni ion in a
cluster with $T_\text{C}=\theta$ is $F(\theta),$ the residual magnetization
$M_{0}(T)$ can be calculated by a simple formula:

\begin{equation} \label{eq_m0}
    M_{0}(T)=m_{0}\int_{T}^{\infty}F(T_\text{C})\sqrt{1-\frac{T^{2}}{T_\text{C}^{2}}} \: dT_\text{C},
\end{equation}

where $m_{0}$ represents
the average concentration of ferromagnetic Ni atoms and their average magnetic
moment. We first assumed the simplest possible scenario, $F(T_\text{C})=\theta^{-1}\exp
(-T_\text{C}/\theta),$ $i.e.,$ an exponential distribution with the width $\theta.$
We can then fit all experimental data very well except for the lowest temperature (Fig.~\ref{fig_mag}, green solid line, $m_{0}=1.65\times10^{-4}$
$\mu_\text{B}$/Ni, $\theta=300$\,K). The deviation of the lowest temperature points from this distribution, suggests that a large number of clusters
have very small Curie temperatures. Indeed, it has been suggested\cite{He2013}, that the critical temperature of Ni
nanoclusters goes down precipitously when the cluster size becomes smaller
than a few nm, and drops to zero with clusters of the order \SIrange{1}{1.5}{\nano\meter}. To
account for this effect we have added a sharp Gaussian to the assumed
distribution function, namely $F(T_\text{C})=\exp(-T_\text{C}/\theta)+4.75\exp
[-(T_\text{C}/\theta^{\prime})^{2}],$ where $\theta^{\prime}=\theta/10$. The
physical meaning is that approximately \SI{30}{\percent} of all ferromagnetic Ni form
ultra-small clusters of only a few nanometers in size, which have a Curie temperature of less
that \SI{30}{\kelvin}. This modified distribution fits the entire range perfectly, as can be seen by the red curve in Fig.~\ref{fig_mag}.

The slope of the $M(H)$ isotherms involved in the extrapolation of the spontaneous moment $M_{0}$ gives the effective susceptibility $\chi_{0}$, which is plotted in the inset to Fig.~\ref{fig_susT}. Within the accuracy of our measurement and analysis, $\chi_{0}$ approaches a temperature independent value of $\sim 2.1\times 10^{-4}$\,emu/mol below 200\,K. We take this value as an estimate of the Pauli susceptibility that would be found in the absence of any ferromagnetic contamination. Using the calculated density of states, which is \SI{4.0}{states\per\electronvolt\per \text{f.u.}}, we compute a bare Pauli susceptibility $\chi_{\mathrm{Pauli}}=1.3\times10^{-4}$\,emu/mol, which when compared with the experimental value of $\chi_{0}$ indicates that there is a moderate magnetic enhancement of the Pauli susceptibility in La$_{2}$Ni$_{2}$In. The measured $\chi_{0}$ agrees rather well with calculations of the Stoner-renormalized spin susceptibility, which appear to be about \SI{33}{\percent} higher than the calculated Pauli susceptibility, i.\,e. 1.7 $\times$10$^{-4}$\,emu/mol. However, note that in itinerant systems, the mean-field DFT calculations tends to \textit{over}estimate, not \textit{under}estimate $\chi$ \cite{moriya}.

\section{Superconducting properties}

\subsection{Specific Heat}

\begin{figure*}[ptb]
\includegraphics[width=\linewidth]{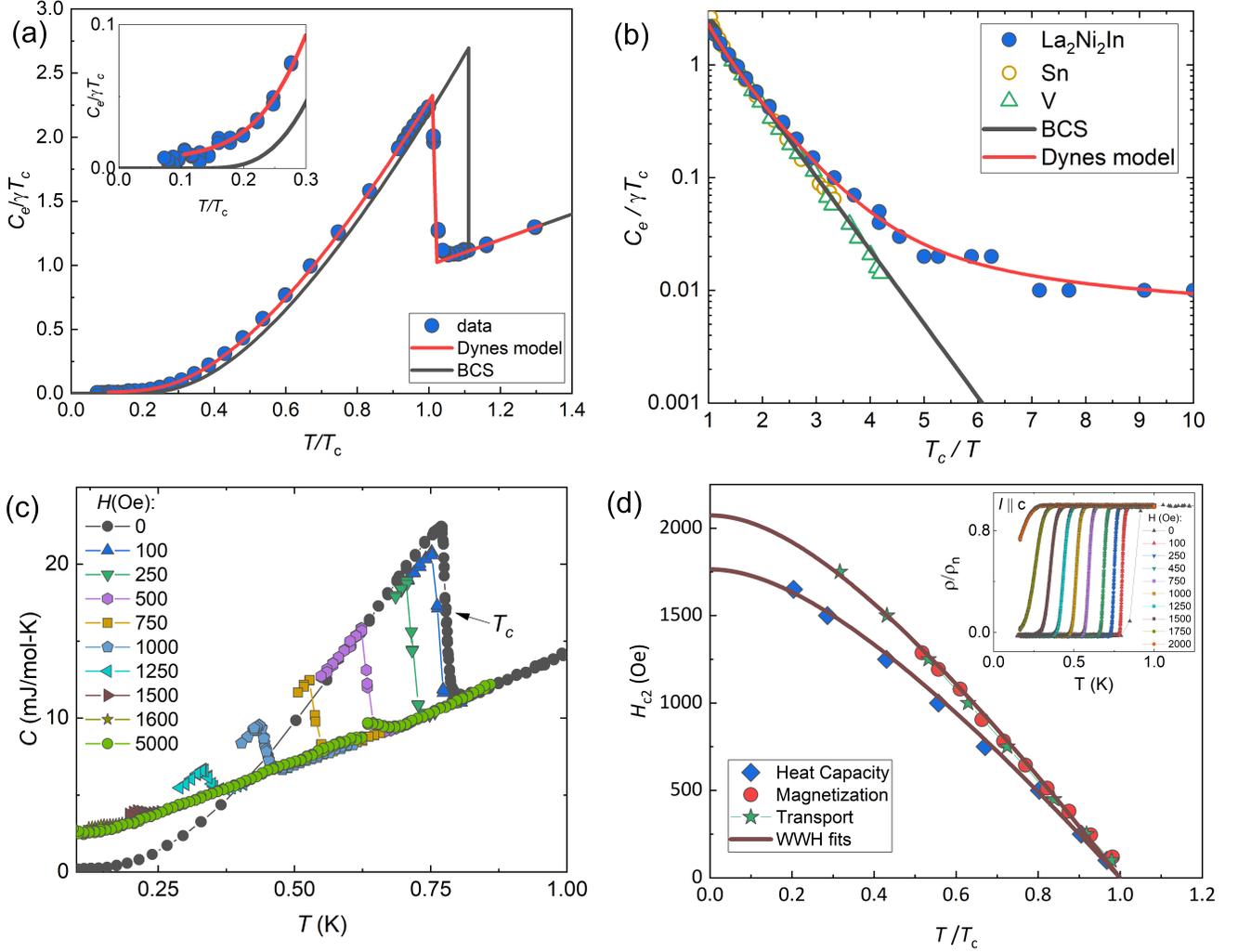}
\caption{Temperature dependence of the specific heat of La$_{2}$Ni$_{2}$In below 1.3\,K. (a) The electronic part of the specific heat is normalized by the product of the Sommerfeld coefficient $\gamma$ and the superconducting transition temperature T$_{c}$ $C_\text{e}/\gamma T_\text{c}$, and is plotted as a function of $T/T_\text{c}$. The inset shows an enlarged view of the low temperature region. Red line indicates fit by Dynes model, described in the text. (b) Semi-log plot of $C_\text{e}/\gamma T_\text{c}$ as a function of the normalized inverse temperature $T_\text{c}/T$ of La$_{2}$Ni$_{2}$In as well as the two BCS superconductors Sn and V for comparison\cite{Hunklinger2018}. Black line is the BCS model, red line is the Dynes model. (c) The suppression of the superconducting jump at T$_{c}$ in the specific heat $C$ is measured with the magnetic field $H\perp c$. Values of the field are as indicated. (d) Derived $H_\text{c2}$-$(T/T_\text{c})$ phase diagram, combining values of T$_{c}$ from specific heat, magnetization, and resistivity measurements. The inset shows the temperature dependencies of the electrical resistivity, measured at different fixed fields. Solid lines represent the WWH model, as discussed in the main text.}%
\label{fig_hc2}%
\end{figure*}

We begin our analysis of the superconductivity in La$_{2}$Ni$_{2}$In with the specific heat measurements. The contribution of the nuclear Schottky effect to the low temperature specific heat $C_{\text{nuc}}$ is described in Appendix~\ref{appendix_nuclear}. The phonon contribution $C_\text{ph}$ was determined by fitting the measured data $C$ between \SIrange{1}{4}{\kelvin} to the expression: $C(T)=\gamma T + \beta_1 T^{3}+\beta_2 T^{5}$. 
The electronic contribution to the specific heat $C_{\text{e}}\ =C-C_{\text{nuc}}-C_{\text{ph}}$ divided by the product of the normal state
Sommerfeld coefficient $\gamma$ and the critical temperature
$T_{\text{c}}$ is plotted as a function of the normalized temperature $T/T_{\text{c}}$ in Figure~\ref{fig_hc2}(a).
The normal state $C_{\text{e}}$ decreases linearly with $T$ as expected for a
conventional metal, until a jump in $C(T)$ at $T=T_c$ occurs that signals a second order phase transition into the SC state. The transition is very
narrow, as previously indicated, confirming bulk SC and a high degree of sample quality.

The magnitude of the specific heat jump
$\Delta C/C_{\text{n}}\approx 1.21 $ is
about \SI{15}{\percent} smaller than the value of 1.43 given by the BCS model in the
weak-coupling limit, with $C_\text{n}$ the normal state specific heat at the transition. In order to better understand this behavior and as well
the evolution of $C_{\text{e}}$ below T$_{\text{c}}$, we examined the dependence of $C_\text{e}/\gamma T_c$, plotted as a function of inverse temperature (Fig.~\ref{fig_hc2}b). While the behavior of $C_{\text{e}}(T)$ in the vicinity of $T_c$ and down to $\approx T_c/4$ is well described by the exponential function predicted by the BCS model, significant deviations are observed at lower temperatures $T<T_c/6$.

The non-exponential temperature dependence of $C_\text{e}(T)$ that is highlighted in Fig.~\ref{fig_hc2}(b) can be described by a sum of an exponential and a linear residual term, suggesting that $C_\text{e}(T)$ in our crystals is affected by the presence of quasiparticles with energies $E<\Delta$. This can indicate either an unconventional pairing symmetry with a nodal order parameter or subgap states in an s-wave superconductor. The subgap states at $E<\Delta$ are not accounted for in the BCS model, in which $C_\text{e}(T)$ is given by:

\begin{equation}
C_\text{e}(T)=\frac{1}{2T^{2}}\int_{0}^{\infty}\biggl[E^{2}-\frac{T}{2}\frac{\partial\Delta^{2}}{\partial T}\biggr]\frac{N(E)dE}{\cosh^{2}(E/2T)}.
\label{C}
\end{equation}
Here $N(E)=N_0 E/\sqrt{E^2-\Delta^2}$ at $E>\Delta$ and $N(E)=0$ at $E<\Delta$ is the BCS density of states, and $N_0$ is the DOS per spin at the Fermi surface in the normal state. Because $N(E)$ vanishes at $E<\Delta$, Eq.~(\ref{C}) gives an exponential temperature dependence of $C_\text{e}^\text{BCS}(T)\sim \gamma T_c\sqrt{\Delta/T}\exp(-\Delta/T)$ at $T\ll \Delta$.

In many materials, the BCS gap singularities in $N(E)$ at $E=\Delta$ are smeared out. The resulting quasiparticle subgap states occurring at $E<\Delta$ have traditionally been addressed in the literature using the phenomenological Dynes model in which\cite{Dynes,JZ}:
\begin{equation}
N(E)=N_0\mbox{Re}\frac{E-i\Gamma}{\sqrt{(E-i\Gamma)^{2}-\Delta^2}}.
\label{N}
\end{equation}
Here a pairbreaking parameter $\Gamma$ accounts for a finite quasiparticle lifetime $\hbar/\Gamma$, resulting in a finite DOS $N(0)=N_0\Gamma/\Delta$ at $E=0$.

To understand the features of $C_\text{e}(T)$ observed in our crystals, we use the Dynes model in which $T_c$, $\Delta(T)$ and $C_\text{e}(T)$ are determined by Eqs.~\ref{tc}-\ref{fs} given in Appendix~\ref{appendix:Dynes}\cite{gk,Herman2017,*Herman2018}. By numerically solving these equations, it is possible to fit the experimental $C_\text{e}(T)$ data, as demonstrated in Fig.~\ref{fig_hc2}(a, b). The Dynes model effectively captures both a non-exponential residual $C_\text{e}(T)$ at low $T$ and the reduction of $\Delta C\approx 1.2 \gamma T_c$ at $T=T_c$, substantially different from the BCS value $\Delta C=1.43\gamma T_c$. The fit was carried out for a dimensionless pairbreaking parameter $g=\Gamma/2\pi T_{c0}= 0.02$, where $\Gamma$ is taken to be independent of $T$. Here $\Gamma$ is about 7 times smaller than the critical value $\Gamma_c=\Delta_0/2$ at which $T_c$ vanishes in the Dynes model (see Appendix~\ref{appendix:Dynes}). In accordance with Fig.~\ref{fig_hc2}, weak pairbreaking at $g=0.02$ produces a small reduction of $T_c$ by about \SI{10}{\percent}, as compared to $T_\text{c0}$ at $\Gamma=0$. This suggests that $T_\text{c0}$ of our crystals would be close to \SI{0.86}{\kelvin} in the ideal case of $\Gamma=0$. The overall expression of $C_\text{e}(T)$ that is calculated from Eq.~(\ref{C}) using the Dynes DOS with $g$=0.02 overall agrees very well with the experimental data.

Subgap states have been revealed by numerous tunneling experiments (see e.g., a review \cite{JZ} and the references therein). Many mechanisms of subgap states have been suggested in the literature, including inelastic scattering of quasiparticles by phonons \cite{inelast}, Coulomb correlations \cite{coulomb}, anisotropy of the Fermi surface \cite{anis}, inhomogeneities of the BCS pairing constant \cite{larkin}, magnetic impurities \cite{Balatskii}, spatial correlations in impurity scattering \cite{Balatskii,meyer}, or diffusive surface scattering\cite{arnold} (see, e.g., Ref. \onlinecite{feigel} for an overview of different mechanisms). The weak ferromagnetism associated with magnetic nanoclusters in La$_{2}$Ni$_{2}$In could potentially affect the value of $\Gamma$ in different ways. We could expect a significant contribution to $\Gamma$ from spin flip magnetic scattering, and as well the presence of localized states associated with magnetic impurities \cite{Balatskii}. Magnetic nanoclusters can also cause local variations of the BCS pairing constant, resulting in a slight broadening of the sharp transition characteristic of the BCS and Dynes models, and as well an additional contribution to $\Gamma$ ~\cite{larkin}. Irrespective of the mechanisms, this analysis gives insight into how an underlying broadening of the DOS gap peaks can account for the behavior of $C_\text{e}(T)$ observed in our crystals. We note that $C_\text{e}(T)$ at $T\lesssim T_c/2$ is mostly determined by thermally-activated quasiparticles with energies $E\approx \Delta$, but at lower temperatures $C_\text{e}(T)$ is dominated by
quasiparticles with $E\sim T\ll \Delta$, leading to a residual specific heat $C_\text{i} \sim \gamma T\Gamma/\Delta$ ~\cite{Herman2017, Herman2018}.

Beyond the effects of magnetic pairbreaking by dilute magnetic impurities~\cite{Openov2004,Maki}, our fits based on the Dynes model accurately describe the jump in the specific heat $\Delta C$ at $T=T_c$ and the overall temperature dependence of $C_\text{e}(T)$ within the superconducting state, while implying that $T_c$ is further reduced from the value determined by only magnetic pairbreaking. We will consider below the possibility that this additional density of states is related to a nodal order parameter in La$_{2}$Ni$_{2}$In.

\subsection{Magnetism and the Superconducting State}

\begin{figure*}[ptb]
\includegraphics[width=\linewidth]{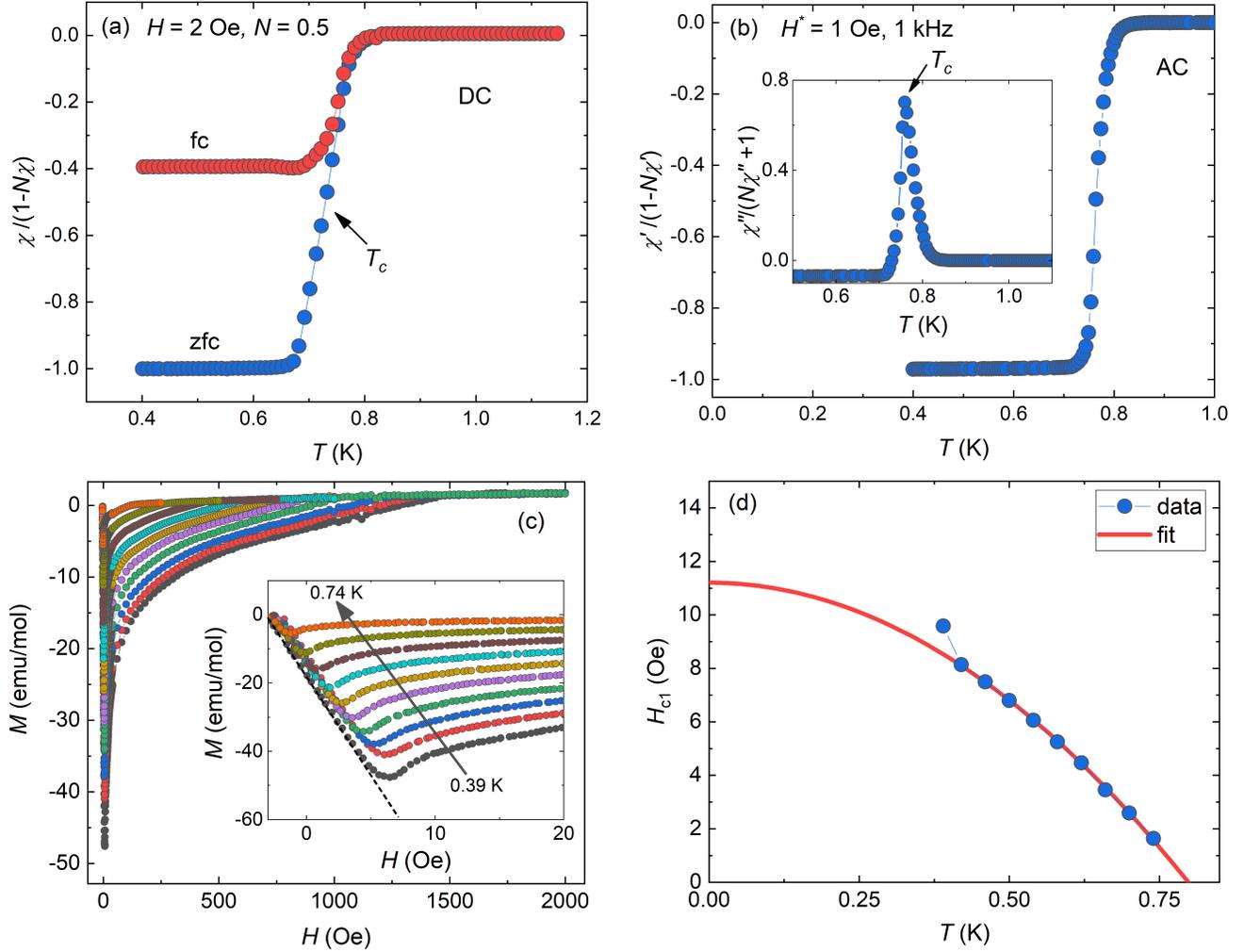}
\caption{Measurements of the magnetization of La$_{2}$Ni$_{2}$In in the superconducting state. (a) Temperature dependencies of the DC field cooled (fc) and
zero-field cooled (zfc) magnetic susceptibility near the superconducting
transition temperature $T_{\text{c}}$. (b) Real(main figure) and imaginary (inset) parts of
the AC susceptibility $\chi^{\prime}$ below 1\,K, with field amplitude $H$* = 1\,Oe and drive
frequency 1\,kHz. (c) Field dependencies of the magnetization measured at fixed temperatures ranging from 0.39\,K to 0.74\,K. The inset shows an
enlarged view of the low field data. The black dashed line is a guide to the
eye that emphasizes the linear part of $M(H)$. (d) Extracted values of the lower critical field presented as a function of temperature
$T$ (solid symbols), including a fit to a Ginzburg-Landau expression (solid line) that is described in the
main text.}
\label{fig_sus}
\end{figure*}

Fig.~\,\ref{fig_sus} shows representative magnetization measurements of La$_{2}$Ni$_{2}$In at temperatures ranging from 0.39\,K to 1.15\,K. The data in Fig.~\,\ref{fig_sus}(a,b) were corrected for demagnetization effects due to the sample shape using the expression for a rectangular cuboid in Ref.~\onlinecite{Prozorov2018}. Bulk superconductivity is evident from the sharp increase of the magnetic susceptibility after zero-field cooling (zfc) as depicted in Fig.~\ref{fig_sus}(a). The superconducting (SC) region is characterized by a SC volume fraction of \SI{99}{\percent}, indicating virtually perfect shielding. The sample also exhibits substantial shielding even after cooling in field (fc), indicating that pinning of flux vortices is small. The corresponding value of the Meissner fraction is \SI{55}{\percent}. Figure~\ref{fig_sus}(b) shows the AC susceptibility recorded with a field amplitude of \SI{1}{Oe} and a drive frequency of \SI{1}{\kilo\hertz}. A sharp peak is visible in the imaginary part of the susceptibility $\chi^{\prime\prime}$, corresponding to the SC transition. The transition temperature, defined as the maximum in $\chi^{\prime\prime}$, was determined to be $T_\text{c} = \SI{0.79}{\kelvin}$. Measurements at various drive frequencies (not shown) found no frequency dependence of the AC susceptibility.

The magnetization isotherms in the temperature range \SIrange{0.39}{0.74}{\kelvin} shown in Figure~\ref{fig_sus}(c) reveal La$_{2}$Ni$_{2}$In to be a type-II superconductor, as evidenced by the linear shielding at low fields (see dashed line in inset). Above about \SI{10}{Oe} the shielding reduces as magnetic flux starts to penetrate the sample and the system enters into the vortex state. Extracting the lower critical field $H_\text{c1}$ from a linear fit to the low-field magnetization results in the phase diagram shown in Fig.~\ref{fig_sus}(d). Its temperature dependence is well described by the Ginzburg-Landau expression: $H_\text{c1}(T) = H_\text{c1}(0) \left(1-(T/T_\text{c})^2\right)$, resulting in a rather small lower critical field value of $H_\text{c1}(0)= \SI{11.2(1)}{Oe}$. However, this values does not account for demagnetization effects. The corrected value is $H_\text{c1}(0)/(1-N) = \SI{14.0(1)}{Oe}$, with $N=0.2$ for the sample used in Fig.~\ref{fig_sus}(c,d).

Upon application of a magnetic field the SC transition is continuously
suppressed to lower temperatures, as shown in Fig.~\ref{fig_hc2}(c). For
fields above \SI{500}{Oe}, SC is fully suppressed and normal metallic behavior is
recovered over the depicted temperature range. Extracting transition temperatures from the inflection points of the curves
and plotting them as a function of the reduced temperature results in
Fig.~\ref{fig_hc2}(d), where we also included data obtained from measurements
of the magnetization and resistivity at various applied fields (see
inset). The $H_{c2}$ data are described well by the single-band Werthamer, Helfand and Hohenberg (WHH) theory \cite{Werthamer1966}. The fit to the data yields an average upper critical field of $H_{c2}(0)=1918(19)$\,Oe at $T=0$. Hence, we can evaluate the Ginzburg-Landau (GL) coherence length $\xi(T)=\xi(0)(1-T/T_c)^{-1/2}$ using the WHH relation $H_{c2}(0)=0.69\phi_0/2\pi\xi^2(0)$, where $\phi_0$ is the magnetic flux quantum. This yields $\xi(0) = 34.4(2)$\,nm.

It is instructive to compare the $T=0$ values of the GL coherence length $\xi(0)$ and the BCS coherence length $\xi_0 = 0.12\hbar v_\text{F}/k_\text{B}T_\text{c}$ at $T=0$, to determine whether La$_{2}$Ni$_{2}$In is in the clean limit \cite{Tinkham2004}. Using the calculated $v_\text{F}\approx 0.2\times 10^8$\,cm/s and $T_\text{c0} = 0.86$\,K from the specific heat data, we get $\xi_0\approx 245$\,nm, about 7 times larger than $\xi(0)$. Such a large difference between $\xi(0)$ and $\xi_0$ indicates that our La$_2$Ni$_2$In crystals are in the dirty limit, where the mean free path due to nonmagnetic impurities $\ell_\text{mfp}$ is much shorter than $\xi_0$. Indeed, an estimate of $\ell_\text{mfp}$ from the residual resistivity and the calculated plasma frequency and Fermi velocity gives $\ell_\text{mfp}\approx 10$\,nm and $\ell_\text{mfp}/\xi_0\simeq 0.04$. The evaluation of the GL coherence length in the dirty limit $\xi(0)=0.855(\xi_0 \ell_\text{mfp})^{1/2}$ ~ \cite{Tinkham2004} yields $\xi(0)\approx 42.3$\,nm, about 24\,\% larger than $\xi(0)$ extracted from the $H_{c2}$ data. Thus, the conclusion that our samples are in the dirty limit is qualitatively consistent with our $H_{c2}$ and transport measurements. This value should be considered to be an upper bound, since a more accurate evaluation of $\xi(0)$ requires taking into account scattering from magnetic impurities, subgap states and strong coupling corrections.

Next, we analyze the lower critical field $H_{\text{c1}}$ using the GL relation:
\begin{equation}
H_{\text{c1}}=\frac{\phi_0}{4\pi\lambda_{\text{L}}^{2}}
\left(\ln
\frac{\lambda_\text{L}}{\xi}  +\frac{1}{2}\right),
\label{hc1}
\end{equation}
where $\lambda_\text{L}$ is the London penetration depth and the factor of 1/2 accounts for the vortex cores.
Evaluation of $\lambda_\text{L0}=\hbar c/\omega_\text{p}$ in the clean limit $\ell_\text{mfp}\gg \xi_0$ gives $\lambda_{L0}\approx 50$\,nm for the field along the c axis.
Using the more appropriate dirty limit expression, $\lambda_\text{L}=\lambda_{L0}\sqrt{1+\xi_0/\ell_\text{mfp}}=250$\,nm in Eq.~(\ref{hc1}), leads to $H_{\text{c1}}\approx 13.7$\,Oe, which is quite close to the experimental value of $14.0$\,Oe. Corrections to $\lambda_{\text{L}}$ due to magnetic impurities and subgap states were estimated to be no more than $\approx 6 \%$, see Appendix~\ref{appendix:Dynes}).

 The values $\xi(0) = 34.4$\,nm and $\lambda_{\text{L}}(0)=250$\,nm were used to obtain the GL parameter $\kappa_{\text{GL}}=\lambda_{\text{L}}/\xi\approx 7.3$, providing additional confirmation that La$_{2}$Ni$_{2}$In is a type-II superconductor. These estimates suggest that if a sample of La$_{2}$Ni$_{2}$In were available without impurities, it would be a type-I superconductor, since $\xi_0=245$\,nm and $\lambda_\text{L}=50$\,nm lead to a value of $\kappa_\text{GL}\approx 0.2$ that is much smaller than the critical GL value 1/$\sqrt{2}$. The putative transition from type-I to type-II superconductor induced by impurities is hardly surprising, given the large BCS coherence length $\xi_0\approx $ 245\,nm, and the low T$_\text{c}$ of La$_{2}$Ni$_{2}$In. Strictly speaking, we cannot rule out the possibility that pristine La$_{2}$Ni$_{2}$In could exhibit unconventional pairing symmetries associated with sign changes in the order parameter. However, the strong impurity scattering in our samples would effectively suppress any nodal states, were they present. For this reason, we used the model of an s-wave superconductor with subgap Dynes states in our analysis of the specific heat data.

\begin{table}[ptb]
\caption{Normal and superconducting properties of La$_{2}$Ni$_{2}$In. The abbreviations "calc" and "exp" attached to some of the variables distinguishes theoretical from experimental results.}
\label{tab_properties}
\begin{ruledtabular}
		\begin{tabular}{lll}
			Parameter & Unit & Value\\
			\hline
			& 		  &				  \\ \vspace{1.5mm}
			$T_\text{c}$ & \si{\kelvin} & 0.89\\ \vspace{1.5mm}
			$H_\text{c2}(0)$ & Oe & 1918 \\ \vspace{1.5mm}
			$\xi(0)$ & \si{\nano\meter} & 34.4 \\ \vspace{1.5mm}
			$\xi^\text{calc}_0$ & \si{\nano\meter} & 245 \\ \vspace{1.5mm}
			$\xi^\text{calc}(0)$ & \si{\nano\meter} & 42.3 \\ \vspace{1.5mm}
			$\ell_\text{mfp}$ & \si{\nano\meter} & 10 \\ \vspace{1.5mm}
			$H_\text{c1}(0)/(1-N)$ & \si{Oe} & 14.0 \\ \vspace{1.5mm}	
			$H^\text{calc}_\text{c1}$ & \si{Oe} & 13.7 \\ \vspace{1.5mm}
			$\lambda^\text{calc}_\text{L}$  &  \si{\nano\meter} & 250 \\ \vspace{1.5mm}
			$\lambda^\text{calc}_\text{L0}$  &  \si{\nano\meter} & 50 \\ \vspace{1.5mm}
			$\kappa_\text{GL}$ & & 7.3 \\ \vspace{1.5mm}
			$\gamma_\text{n}$ & \si{\milli\joule\per\mole\per\kelvin\tothe{2}}& 13.01 \\ \vspace{1.5mm}
			$\beta$ & \si{\milli\joule\per\mole\per\kelvin\tothe{4}} & 0.84 \\ \vspace{1.5mm}
			$T_\text{D}$ & \si{\kelvin}&  226 \\ \vspace{1.5mm}
			$\lambda_\text{MM}$ & &  0.461 \\ \vspace{1.5mm}
			$\lambda_\text{tr}$ & &  0.395 \\ \vspace{1.5mm}
			$\lambda_{m^\ast/m}$ & &  0.38 \\ \vspace{1.5mm}
			$\omega_\text{p}^\text{a}$ & eV & 3.31 \\ \vspace{1.5mm}
			$\omega_\text{p}^\text{c}$ & eV & 3.98 \\ \vspace{1.5mm}
			$N(0)$ & \si{states\per\electronvolt\per\text{f.u.}} &   3.96\\ \vspace{1.5mm}
			$\chi_\text{Pauli}$ & $10^{-4}$\si{\text{emu/mol}} &   1.3\\ \vspace{1.5mm}
		$\chi_\text{calc}$ & $10^{-4}$\si{\text{emu/mol}} &   1.7\\ \vspace{1.5mm}
		$\chi_\text{exp}$ & $10^{-4}$\si{\text{emu/mol}} &   2.1\\ \vspace{1.5mm} 
			$\Delta C/\gamma_\text{n}T_\text{c}$ & & 1.21 \\ \vspace{1.5mm}
			$\Delta(0)/k_\text{B}T_\text{c}$ & & 1.26 \\
		\end{tabular}
	\end{ruledtabular}
\end{table}

The results for the Debye temperature $T_{\text{D}}\ =\SI{226}{\kelvin}$ and
critical temperature $T_{\text{c0}}\ =\SI{0.86}{\kelvin}$ allow us to estimate
the strength of the electron-phonon coupling from the
McMillan equation\cite{McMillan1968}:
\[
\lambda_{\text{MM}}\ =\frac{1.04+\mu^{\ast}\ln(T_{D}/1.45T_{\text{c0}}%
)}{(1-0.62\mu^{\ast})\ln(T_{D}/1.45T_{\text{c0}})-1.04},
\]
with the Coulomb repulsion parameter assumed to be $\mu^{\ast}=0.13.$ We get a
value of $\lambda_{\text{MM}}=0.46$, reasonably close to our previous estimates above. This indicates that La$_{2}$Ni$_{2}%
$In is a superconductor in the weak to intermediate
coupling regime. The value derived above is in agreement with the mass renormalization deduced from specific heat and from $\lambda_\text{tr}$, which are 0.38 and 0.395, respectively. A slightly larger value of the McMillan $\lambda_\text{MM}$ likely reflects the difference between the Debye and the logarithmic frequencies.

A summary of the various properties derived in this analysis can be found in Table~\ref{tab_properties}.

\section{Conclusion}

We have, for the first time, synthesized high-quality single crystals of La$_{2}$Ni$_{2}$In, previously only available in polycrystalline form. Resistivity measurements show good metallic behavior, in agreement with DFT calculations. The density of states taken from the Sommerfeld coefficient of the specific heat is only slightly enhanced relative to the one determined from the DFT calculations, signalling only weak electronic correlations. Unlike the more highly studied members of the $R_{2}T_{2}X$ ($R$=rare earth, $T$=transition metal, $X$= main group element) family, DFT calculations indicate that the Ni states lie well below the Fermi energy, with a substantial degree of charge transfer that ensures that the Ni magnetism is quenched. Measurements of the magnetization reveal weak ferromagnetism that is associated with ferromagnetic contamination, most likely elemental Ni. The remainder of the magnetic susceptibility is nearly temperature independent, as expected for the Pauli susceptibility. Relative to the value of the Pauli susceptibility expected for the density of states taken from the measured Sommerfeld constant, we infer that there is a weak enhancement that is of similar magnitude to the small Stoner factor determined from the DFT calculations. Our measurements and the DFT calculations together imply that La$_{2}$Ni$_{2}$In is best understood as a good metal with minimal electronic correlations. Superconductivity is observed below 0.9\,K, in good agreement with the McMillan expression for $T_{c}$, using values of the electron-phonon interaction taken from the DFT calculations. A detailed analysis of the magnetic susceptibility and specific heat in the superconducting state find that La$_{2}$Ni$_{2}$In is a type-II superconductor that is in the dirty limit, although there are indications that it could be type-I in the absence of impurities. Our analysis using the phenomenological Dynes model highlights important roles for subgap quasiparticle states, beyond those expected from pairbreaking alone. To our knowledge, La$_{2}$Ni$_{2}$In is the first superconductor reported in this family of compounds. The weakness of the Ni magnetism, and the absence of magnetic correlations are likely to make it a conventional superconductor. Considering the spectrum of behaviors that have already been observed in f-electron bearing members of this family of compounds, which range from strong local-moment magnetism, to mixed valence, and ultimately to conventional metals with differing degrees of correlations, we place La$_{2}$Ni$_{2}$In in this last category. In this way, it should be considered as analogous to other conventional superconductors with nonmagnetic or weakly magnetic Ni. Using the Sommerfeld coefficient as a proxy for the strength of electronic correlations, we find that La$_{2}$Ni$_{2}$In with $\gamma
=\SI{13}{\milli\joule\per\mole\per\kelvin\tothe{2}}$ and $T_\text{c}$=0.89\,K is much more strongly correlated than LaNiAsO ($\gamma
=\SI{3}{\milli\joule\per\mole\per\kelvin\tothe{2}}$ and $T_\text{c}$=2.7\,K)\cite{Li2014}, but not as correlated as La$_{3}$Ni ($\gamma
=\SI{21}{\milli\joule\per\mole\per\kelvin\tothe{2}}$ and $T_\text{c}$=2.2\,K)\cite{Sato1994}, and the most correlated La$_{7}$Ni$_{3}$ $\gamma
=\SI{44}{\milli\joule\per\mole\per\kelvin\tothe{2}}$ and $T_\text{c}$=2.4\,K)\cite{Nakamura2017}.

\section{Acknowledgments}
This research was supported by NSF-DMR-1807451. JM was supported in part by funding from the Max Planck-UBC-UTokyo Centre for Quantum Materials and the Canada First Research Excellence Fund, Quantum Materials and Future Technologies Program.

\bibliography{references}

\appendix

\section{Comparison of the Tetragonal and Orthorhombic Structure Variants}\label{sec_ortho}

\begin{figure}
\includegraphics[width=\linewidth]{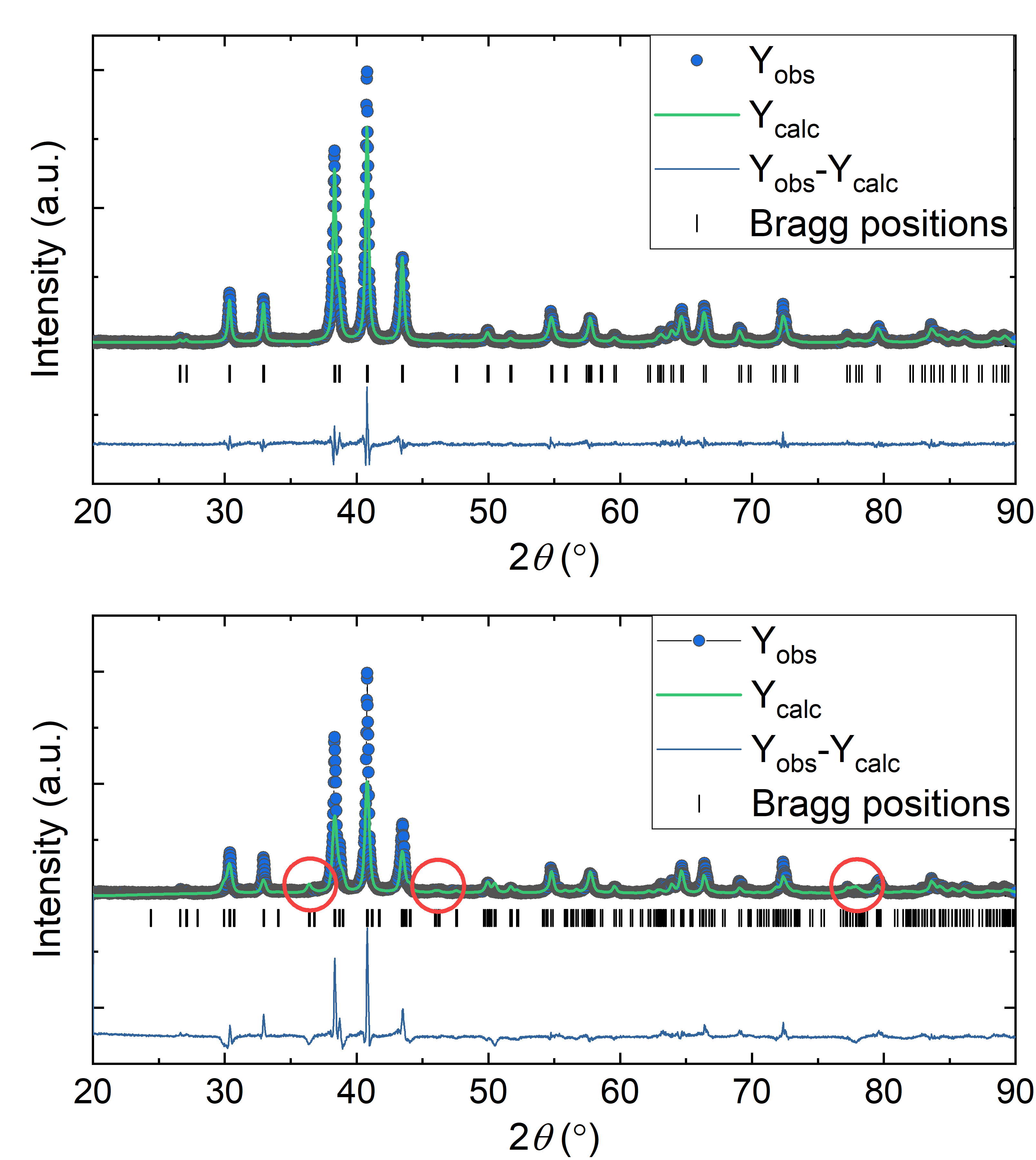}
\caption{Comparison between Rietveld refinements of the XRD data of La$_2$Ni$_2$In using the the tetragonal model with space group P4/mbm (top) and the forced orthorhombic structure with space group Pbam. The orthorhombic model does not fit the data as well as the tetragonal model. In particular, additional peaks are present in the orthorhombic model (circles) that are not observed in the data.}%
\label{fig_supp_xrd}%
\end{figure}

Pustovoychenko et al. recently reported on the synthesis of an orthorhombic variant of La$_2$Ni$_2$In in Ref.~\onlinecite{Pustovoychenko2012}. Since our DFT calculations also seemed to favor this structure, we forced a fit with the orthorhombic Pbam structure to our recorded XRD data. A direct comparison between the data fitted to the tetragonal and orthorhombic structure, respectively, can be seen in Fig.~\ref{fig_supp_xrd}. Since the orthorhombic structure features peaks, that are not present in our data, we conclude that we have grown the tetragonal variant.

\section{Density functional calculations of the phonons} \label{appendix_dft}

Motivated by the deviation of the specific heat to the standard Debye model we calculated the zone-center phonon spectrum. The respective energies are listed in Table~\ref{tab_phonons}.

\begin{table}
  \centering
  \caption{Calculated phonon energies of La$_2$Ni$_2$In at the center ($\Gamma$-point) of the Brillouin zone.}
  \label{tab_phonons}
\begin{ruledtabular}
    \begin{tabular}{rrrrr}
        \#  & \multicolumn{4}{c}{\centering Phonon Energy} \\
    \multicolumn{1}{l}{} & \multicolumn{1}{c}{[THz]} & \multicolumn{1}{c}{[2$\pi$THz]} & \multicolumn{1}{c}{[cm$^{-1}$]} & \multicolumn{1}{c}{[meV]}\vspace{1mm}\\
    \hline
          &         &          &          & \\
    1     & 4.22229 & 26.52944 & 140.8405 & 17.462 \\
    2     & 4.20341 & 26.41078 & 140.2106 & 17.3839 \\
    3     & 4.00624 & 25.17197 & 133.6339 & 16.5685 \\
    4     & 4.00624 & 25.17197 & 133.6339 & 16.5685 \\
    5     & 3.78668 & 23.79244 & 126.3102 & 15.66047 \\
    6     & 3.54148 & 22.25177 & 118.131 & 14.64638 \\
    7     & 3.54148 & 22.25177 & 118.131 & 14.64638 \\
    8     & 3.52643 & 22.15721 & 117.629 & 14.58414 \\
    9     & 3.28964 & 20.66944 & 109.7307 & 13.60488 \\
    10    & 3.28964 & 20.66944 & 109.7307 & 13.60488 \\
    11    & 3.07298 & 19.30812 & 102.5037 & 12.70884 \\
    12    & 3.06187 & 19.23831 & 102.1331 & 12.66289 \\
    13    & 2.98852 & 18.7774 & 99.68617 & 12.35952 \\
    14    & 2.80868 & 17.64744 & 93.68741 & 11.61576 \\
    15    & 2.80827 & 17.64488 & 93.67381 & 11.61408 \\
    16    & 2.80827 & 17.64488 & 93.67381 & 11.61408 \\
    17    & 2.70542 & 16.99868 & 90.24323 & 11.18874 \\
    18    & 2.70542 & 16.99868 & 90.24323 & 11.18874 \\
    19    & 2.55974 & 16.08329 & 85.38359 & 10.58622 \\
    20    & 2.41802 & 15.19286 & 80.65641 & 10.00012 \\
    21    & 2.34038 & 14.70504 & 78.06667 & 9.67904 \\
    22    & 2.34038 & 14.70504 & 78.06667 & 9.67904 \\
    23    & 2.11171 & 13.26824 & 70.43892 & 8.73332 \\
    24    & 2.10024 & 13.19619 & 70.05642 & 8.68589 \\
    25    & 1.78281 & 11.20173 & 59.46814 & 7.37311 \\
    26    & 1.53756 & 9.66074 & 51.2873 & 6.35882 \\
    27    & 1.53756 & 9.66074 & 51.2873 & 6.35882 \vspace{1mm} \\
    \end{tabular}%
  \label{tab:phonons}%
  \end{ruledtabular}
  \label{tab_supp_phonons}
\end{table}%

\section{Sample Dependence of the Magnetic Susceptibility} \label{appendix_sample_depend}

\begin{figure*}[ptb]
\centering
\includegraphics[width=\linewidth]{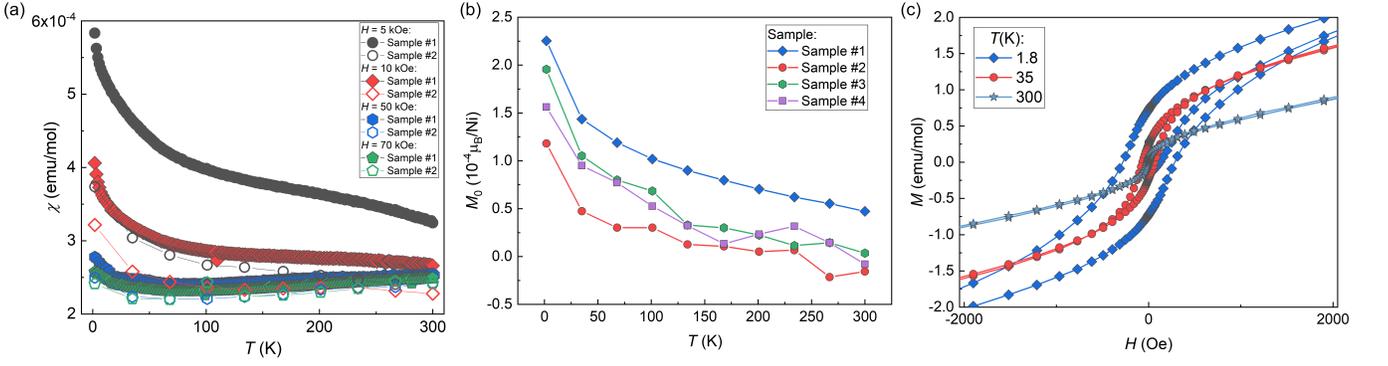}
\caption{Magnetic properties of various single crystals of La$_2$Ni$_2$In from the same batch. (a) Temperature dependencies of the magnetic susceptibility $\chi=M/H$ with measuring field as indicated and applied along the c-axes. (b) Temperature dependencies of the spontaneous magnetization $M_0$ for 4 samples. (c) Full hysteresis loops at 1.8/,K, 35\,K and 300\,K measured on Sample \#1.}%
\label{fig_supp_sus}%
\end{figure*}

The magnetic properties of four samples of La$_2$Ni$_2$In were measured for comparison. Sample \#1 consisted of a co-aligned stack of six single crystals, while Samples \#2-\#4 were single crystals. All crystals were taken from the same batch and the measuring field was always applied along the c-axis. Fig.~\ref{fig_supp_sus}(a) depicts the temperature dependence of the DC magnetic susceptibility $\chi=M/H$ for samples \#1 and \#2, while Fig.~\ref{fig_supp_sus}(b) shows the temperature dependence of the spontaneous magnetization $M_0$ for four samples determined from the high-temperature fit to the magnetization $M(H)$ as discussed in the main text. The temperature dependencies of both $\chi$ and $M_0$ are similar among the samples. In the context of the ferromagnetic cluster model described in the main text, this suggests that there is little variation in the distribution of cluster sizes among the different crystals. However, the magnitudes of $\chi$(T) vary among the crystals by approximately a factor of two, corresponding to a factor of two variation in the overall amount of ferromagnetic clusters that is present in the different crystals. Overall this is a very reasonable result, considering that the origin of the ferromagnetism is most likely the inclusion of unreacted Ni from the LaNi precursor, which is common for all crystals from a single preparation batch.

Finally, Fig.~\ref{fig_supp_sus}(c) depicts the hysteresis present in La$_2$Ni$_2$In at three indicated temperatures. The magnitude of the hysteresis decreases with increasing temperture.

\section{Isolating the Nuclear Specific Heat} \label{appendix_nuclear}

\begin{figure}[ptb]
\centering
\includegraphics[width=\linewidth]{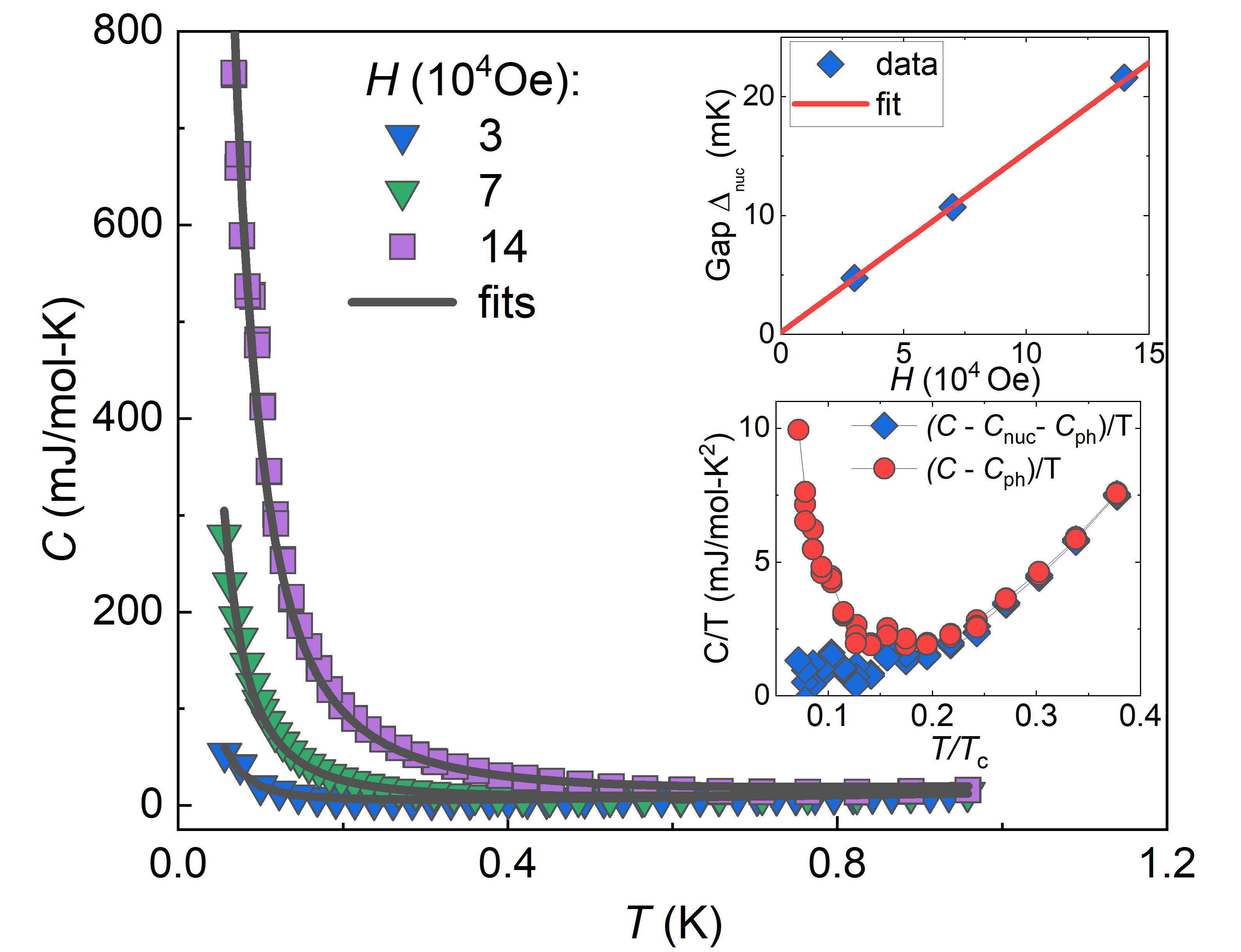}
\caption{Low-temperature specific heat data as a function of temperature $T$
between \SIrange{0.05}{1}{\kelvin}, measured in several magnetic fields. The data were
fit to the model $C(T)=C_\text{nuc}\ +C_\text{e}\ +C_\text{ph}\ $ defined in the
main text. The upper inset shows the extracted nuclear energy gap as a function of
applied magnetic field, together with a linear fit to the data (solid red
line), while the lower inset shows the specific heat over temperature before and after the subtraction of the nuclear term.}%
\label{fig_supp_hc}%
\end{figure}

Both La and In have large nuclear spins of $I = 7/2$ and $I=9/2$, respectively, and thus they may generate nuclear Schottky anomalies in the low temperature specific heat of La$_2$Ni$_2$In. In fact, our measured specific heat does exhibit a sharp increase at low temperatures in a magnetic field (Fig.~\ref{fig_supp_hc}), that gets more pronounced with increasing magnetic field. It seems likely that this contribution to the specific heat is a Schottky anomaly that is related to nuclear energy levels in either the La or In atoms. The 2-level Schottky expression is given by
\[
 C_\text{nuc}(T) = R_0\left(\frac{\Delta_\text{nuc}}{T}\right)^2\frac{\exp(\Delta_\text{nuc}/T)}{\left[1+\exp(\Delta_\text{nuc}/T)\right]^2},
\]
with energy gap $\Delta$ and $R_0$ the universal gas constant. At high temperatures, $T \gg \Delta$, this reduces to:
\[
C_\text{nuc}(T) \sim \frac{R_0}{4}\left(\frac{\Delta_\text{nuc}}{T}\right)^2 = \frac{A}{T^2}.
\]
 The respective least-squares fits are compared to the data in Fig.~\ref{fig_supp_hc}. The upper inset shows the field-dependence of the derived energy gap $\Delta_\text{nuc}$, which increases linearly from $\Delta_\text{nuc}$=\SI{0.17(16)}{\milli\kelvin} at $H$=0, signalling that the nuclear levels undergo a Zeeman splitting in the external magnetic field. Extrapolating to $H=0$, we get a value for the coefficient of $A=\SI{0.00(2)}{\milli\joule\kelvin\per\mole}$. Choosing $A=\SI{1.55}{\mu\joule\kelvin\per\mole}$ within the error bar, yields the corrected data that we used in our analysis. A comparison of the as-measured specific heat $C(T)$ before and after the subtraction of $C_\text{nuc}$(T) is depicted in the lower inset to Fig.~\ref{fig_supp_hc}.

\section{Dynes model}
\label{appendix:Dynes}

Here we present the formulas of the Dynes model ~\cite{gk, Herman2017,Herman2018} used in our fits of $C_\text{e}(T)$. The
equations for $\Delta(T)$ and $T_c$ are:
\begin{gather}
\ln\frac{T}{T_\text{c0}}=2\pi T\sum_{\omega>0}\biggl[\frac{1}{\sqrt{(\omega+\Gamma)^{2}+\Delta^{2}}}-\frac{1}{\omega}\biggr],
\label{del}
\\
\ln\frac{T_\text{C}}{T_\text{c0}}+\psi\left(\frac{1}{2}+\frac{\Gamma}{2\pi T_{c}}\right)-\psi\left(\frac{1}{2}\right)=0,
\label{Tc}
\end{gather}
where $\psi(z)$ is a digamma function, $\omega$ is the Matsubara frequency, and $T_{c0}$ is a critical temperature at $\Gamma=0$. Equation (\ref{Tc}) has the same form as the Abrikosov-Gorkov equation for $T_c$ in superconductors with magnetic impurities, so $T_c$ decreases with $\Gamma$ and vanishes at $\Gamma_c=\Delta_0/2$, where $\Delta_{0}$ is the gap at $T=0$ and $\Gamma=0$ ~\cite{Maki,Balatskii}. At $\Gamma\ll T_c$, Eqs. (\ref{del}) and \ref{Tc}) yield:
\begin{gather}
T_{c}=T_{c0}-\frac{\pi\Gamma}{4},
\label{tcc}
\\
\Delta\simeq\Delta_{0}-\Gamma-\frac{\pi^{2}\Gamma T^2}{6\Delta_0^{2}}, \quad T\ll T_c.
\label{del0}
\end{gather}
 The finite DOS at $E=0$ in the Dynes model results in a quadratic temperature dependence of $\Delta(T)$ instead of the BCS exponential behavior of $\Delta(T)\simeq \Delta_0-\sqrt{2\pi T\Delta_0}\exp(-\Delta_0/T)$ at $T\ll T_{c}$.

The magnetic penetration depth in the dirty limit is~ \cite{gk}:
\begin{equation}
\frac{1}{\lambda_\text{L}^{2}}=\frac{\pi\mu_{0}\Delta}{\hbar\rho_{n}}\mbox{Im}\psi\left(\frac{1}{2}+\frac{\Gamma}{2\pi T}+\frac{i\Delta}{2\pi T}\right),
\label{lam}
\end{equation}
where $\rho_n$ is the normal state resistivity. At $T\ll T_c$ Eq. (\ref{lam}) reduces to:
\begin{equation}
\frac{1}{\lambda_\text{L}^{2}}=\frac{2\mu_{0}\Delta}{\hbar\rho_{n}}\tan^{-1}\frac{\Delta}{\Gamma}
\label{lam0}
\end{equation}

The specific heat $C_\text{e}=-T\partial^2F/\partial T^2$ is calculated using the free energy $F$ in the Dynes model in which $F$ is obtained by substituting $\omega\to\omega+\Gamma$ into the BCS formula for $F$. The result can be written in the form:
\begin{gather}
F=F_n+4\pi T N_0\Delta^2\sum_{\omega>0}\biggl[\frac{1}{2\sqrt{(\omega+\Gamma)^2+\Delta^2}} -
\nonumber \\
-\frac{1}{\omega+\Gamma+\sqrt{(\omega+\Gamma)^2+\Delta^2}}\biggr],
\label{F}
\end{gather}
where $F_n=-\pi^2N_0T^2/3$ is the free energy of the normal state and $N_0$ is the density of states per spin.

It is convenient to recast Eqs. (\ref{del}), (\ref{Tc}) and (\ref{F}) in the dimensionless form:
\begin{gather}
\ln t_c+\psi\left(\frac{1}{2}+\frac{g}{t_c}\right)-\psi\left(\frac{1}{2}\right)=0,
\label{tc}
\\
\!\!\ln t=\sum_{n=0}^{\infty}\biggl[\frac{1}{\sqrt{(n+1/2+g/t)^{2}+s/t^{2}}}-\frac{1}{n+1/2}\biggr],
\label{d}
\end{gather}
where $t=T/T_{c0}$, $t_c=T_c/T_{c0}$, $g=\Gamma/2\pi T_{c0}$, and $s=(\Delta/2\pi T_{c0})^{2}$.
The normalized specific heat $c=C_\text{e}(T)/C_n(T_{c0})$, where $C_n(T_{c0})=2\pi^2 N_0T_{c0}/3$, is then:
\begin{gather}
c(t)=t-t\frac{\partial^2f_s}{\partial t^2},
\label{c} \\
f_s=6s\sum_{n=0}^{\infty}\biggl[\frac{1}{\sqrt{(n+1/2+g/t)^{2}+s/t^{2}}}-
\nonumber \\
\frac{2}{n+1/2+g/t+\sqrt{(n+1/2+g/t)^{2}+s/t^{2}}}\biggr].
\label{fs}
\end{gather}
Equations (\ref{tc})-(\ref{fs}) were solved numerically to fit the experimental data shown in Fig. 7. The fit was done with $g=\Gamma/2\pi T_{c0}=0.02$ and $\Gamma$ independent of $T$. Here $g=0.02$ corresponds to $\Gamma/\Delta_0=2\pi g/1.76=0.07$ about 7 times smaller than $\Gamma_{c}=\Delta_{0}/2$ at which $T_{c}\to0$.
At $g=0.02$ weak pairbreaking results in a small increase of $\lambda_\text{L}\approx 1.06\lambda_\text{L0}$ given by Eq.~(\ref{lam0}).

\end{document}